\documentclass[aps,prc,twocolumn,showpacs,preprintnumbers,nofootinbib,float,superscriptaddress,longbibliography]{revtex4-2}
\usepackage{graphicx}
\usepackage[dvipsnames]{xcolor}
\usepackage[colorlinks=true, pdfstartview=FitV, linkcolor=RedOrange, citecolor=ForestGreen, urlcolor=blue]{hyperref}
\usepackage{amsmath}
\usepackage{bm}

\newcommand{\snn}{\sqrt{s_\mathrm{NN}}}
\newcommand{\pT}{p_\mathrm{T}}
\newcommand{\ybeam}{y_\mathrm{beam}}
\newcommand{\esw}{e_\mathrm{sw}}

\newcommand{\bs}{\boldsymbol}

\begin{document}

\title{Bayesian Model Selection and Uncertainty Propagation for Beam Energy Scan Heavy-Ion Collisions}

\author{Syed Afrid Jahan} \email{hm0746@wayne.edu}
\affiliation{Department of Physics and Astronomy, Wayne State University, Detroit, Michigan 48201, USA}
\author{Hendrik Roch} \email{Hendrik.Roch@wayne.edu}
\affiliation{Department of Physics and Astronomy, Wayne State University, Detroit, Michigan 48201, USA}
\author{Chun Shen} \email{chunshen@wayne.edu}
\affiliation{Department of Physics and Astronomy, Wayne State University, Detroit, Michigan 48201, USA}

\begin{abstract}
We apply the Bayesian model selection method (based on the Bayes factor) to optimize $\sqrt{s_\mathrm{NN}}$-dependence in the phenomenological parameters of the (3+1)-dimensional hybrid framework for describing relativistic heavy-ion collisions within the Beam Energy Scan program at the Relativistic Heavy-Ion Collider. The effects of various experimental measurements on the posterior distribution are investigated. We also make model predictions for longitudinal flow decorrelation, rapidity-dependent anisotropic flow and identified particle $v_0(p_\mathrm{T})$ in Au+Au collisions, as well as anisotropic flow coefficients in small systems. Systematic uncertainties in the model predictions are estimated using the variance of the simulation results with a few parameter sets sampled from the posterior distributions.
\end{abstract}

\maketitle

\section{Introduction}
\label{sec:intro}
Hot and dense QCD matter is studied through high-energy heavy-ion collisions at the Relativistic Heavy-Ion Collider (RHIC) and Large Hadron Collider (LHC)~\cite{Achenbach:2023pba, Arslandok:2023utm, Busza:2018rrf}.
These collisions create a hot, strongly interacting fluid of quarks and gluons, known as the Quark-Gluon Plasma (QGP), offering a unique way to probe QCD matter under extreme conditions~\cite{Heinz:2013th, Gale:2013da}. 
Since theoretical calculations of QGP properties from first principles are extremely challenging~\cite{Moore:2020pfu, Ratti:2018ksb}, comparisons between experimental data and phenomenological models are essential for extracting meaningful insights~\cite{Shen:2020mgh, Heinz:2024jwu}.

Direct measurements of the short-lived, small-scale quark-gluon plasma in relativistic nuclear collisions are impossible. 
Instead, researchers analyze the momentum distributions of final-state particles to infer the properties of the QGP and its dynamical evolution during the collision process. 
To quantitatively interpret these observables, multi-stage phenomenological models are developed to describe different stages of the collisions — from the initial-state and pre-equilibrium phase~\cite{Schlichting:2019abc, Mazeliauskas:2021lqi}, to the relativistic viscous hydrodynamics stage~\cite{Heinz:2009xj, Denicol:2012cn, Chaudhuri:2013yna, Jeon:2015dfa}, to the final particlization~\cite{Cooper:1974mv} and hadronic rescatterings~\cite{Bass:1998ca, Bleicher:1999xi, SMASH:2016zqf}. 
These models provide insights into key properties of QCD matter, such as transport properties, critical points, and phase transitions~\cite{Song:2010mg, florkowski2010phenomenology, Shen:2015msa, Pasechnik:2016wkt, Dubla:2018czx, Du:2024wjm}.

Extracting the properties of QCD matter from these phenomenological models is a general inverse problem, which requires computationally intensive event-by-event simulations~\cite{Jacobs:2025ncn}. 
Bayesian inference analysis provides a systematic method for these inverse problems~\cite{Paquet:2023rfd}. 
In recent years, Bayesian analyses enabled by training fast surrogate emulators, have given strong constraints on the temperature dependence of QGP specific shear and bulk viscosities and parton energy-loss parameter $\hat{q}$~\cite{Bernhard:2015hxa, Bernhard:2016tnd, Bernhard:2019bmu, JETSCAPE:2020mzn, JETSCAPE:2021ehl, Parkkila:2021yha, Xie:2022ght, Liyanage:2023nds, Shen:2023awv, JETSCAPE:2023nuf, Jahan:2024wpj, JETSCAPE:2024cqe, Gotz:2025wnv}. 

Data from the RHIC Beam Energy Scan (BES) program offers a valuable opportunity for studying the QGP properties at finite net baryon densities~\cite{STAR:2017sal, STAR:2016vqt}. In Refs.~\cite{Shen:2023awv, Shen:2023pgb, Roch:2024xhh, Jahan:2024wpj}, we performed systematic Bayesian inference for Au+Au collisions at the RHIC BES energies and extract $\mu_B$-dependent QGP viscosity and initial-state stopping power. In this work, we will explore improvements in model calibration by introducing collision energy dependence for specific parameters. We employ the Bayesian model selection method to judge whether it is worthwhile to introduce additional model parameters. Specifically, we compare the Bayes evidence among different models~\cite{Trotta:2008qt, JETSCAPE:2020mzn, Paquet:2023rfd}, which serves as a statistically robust measure for comparing different models in Bayesian analysis to maximize information gain. With the optimize model selection, we will then extend the scope of our analysis by performing inference with more observables, namely more identified hadron yields, charged hadron $p_\mathrm{T}$ fluctuations~\cite{STAR:2019dow}, and their scalar-product $p_T$-differential elliptic flow, $v_2\{\mathrm{SP}\}(\pT)$~\cite{STAR:2004jwm, STAR:2012och}.

The rest of the paper is structured as follows: Section~\ref{sec:model} outlines the theoretical model used for data simulations and training the surrogate Gaussian Process (GP) emulators. Section~\ref{sec:BMS} discusses the application of Bayesian model selection to optimize model calibration by introducing collision energy-dependent parameters. With the optimized model, we expand our scope and perform inference with more experimental observables in Section~\ref{sec:vary_exp_data}. We will quantify the constraining power of these experimental observables on the model posterior distributions. In Section~\ref{sec:prediction}, we will generate samples from the posterior distributions and provide model predictions for new experimental observables, namely longitudinal flow decorrelation, rapidity-dependent elliptic flow, anisotropic flow in small collision systems, and identified particle $v_0(\pT)$ at the RHIC BES energies.

\section{The Theoretical Framework}
\label{sec:model}
In this work, we use the simulation data for Au+Au collisions at $\snn = 7.7, 19.6$, and $200$\,GeV produced with the \texttt{iEBE-MUSIC} framework~\cite{jahan_2024_12807556, Shen:2022oyg, Shen:2023awv}.
The \texttt{3D-Glauber} model~\cite{Shen:2017bsr, Shen:2022oyg} is used to initialize the (3+1)D dynamical evolution for the collision systems. 
It is coupled to second-order relativistic viscous hydrodynamics (\texttt{MUSIC})~\cite{Schenke:2010nt, Paquet:2015lta, Denicol:2018wdp} solved with a lattice QCD based \texttt{NEOS-BQS} equation of state~\cite{Monnai:2019hkn}, particlization sampler (\texttt{iSS})~\cite{Shen:2014vra} and a hadronic transport model (\texttt{UrQMD})~\cite{Bass:1998ca, Bleicher:1999xi}.
The \texttt{3D-Glauber} model offers a dynamical initialization setup that accounts for the finite thickness of the nuclei along the collision axis as they pass through each other~\cite{Shen:2017ruz, Shen:2017bsr}. 
Such a dynamical initialization scheme is an essential ingredient for simulating heavy-ion collisions with collision energy around $\mathcal{O}(10)$~GeV~\cite{Shen:2017fnn, Shen:2018pty, Shen:2021nbe, Shen:2023aeg}.

The training dataset was generated in a 20-dimensional parameter space with the \texttt{3D-Glauber+MUSIC+UrQMD} model~\cite{Shen:2023awv, Jahan:2024wpj}. Their prior ranges are listed in Table~\ref{tab:parameters}.
\begin{table}[t!]
    \caption{The 20 model parameters and their prior ranges.}
    \label{tab:parameters}
    \begin{tabular}{c|c|c|c}
    \hline\hline
    Parameter & Prior & Parameter & Prior \\
    \hline
    $B_G\;[\mathrm{GeV}^{-2}]$ & $[1,25]$ & $\alpha_{\text{string tilt}}$ & $[0,1]$ \\
    $\alpha_{\rm shadowing}$ & $[0,1]$ & $\alpha_{\text{preFlow}}$ & $[0,2]$ \\
    $y_{{\rm loss},2}$ & $[0,2]$ & $\eta_0$ & $[0.001,0.3]$ \\
    $y_{{\rm loss},4}$ & $[1,3]$ & $\eta_2$ & $[0.001,0.3]$ \\
    $y_{{\rm loss},6}$ & $[1,4]$ & $\eta_4$ & $[0.001,0.3]$ \\
    $\sigma_{y_{\rm loss}}$ & $[0.1,0.8]$ & $\zeta_{\rm max}$ & $[0,0.2]$ \\
    $\alpha_{\rm rem}$ & $[0,1]$ & $T_{\zeta,0}\;[{\rm GeV}]$ & $[0.15,0.25]$\\
    $\lambda_B$ & $[0,1]$ & $\sigma_{\zeta,+}\;[{\rm GeV}]$ & $[0.01,0.15]$ \\
    $\sigma_x^{\rm string}\;[{\rm fm}]$ & $[0.1,0.8]$ & $\sigma_{\zeta,-}\;[{\rm GeV}]$ & $[0.005,0.1]$ \\
    $\sigma_\eta^{\rm string}$ & $[0.1,1]$ & $e_{\rm sw}\;[{\rm GeV}/{\rm fm}^3]$ & $[0.15,0.5]$\\
    \hline\hline
    \end{tabular}
\end{table}

The initial-state model describes how the incoming nuclei lose energy and momentum with the following parameters. Parameters $y_{\rm loss,2, 4, 6}$ model the averaged rapidity loss as a piece-wise function of the incoming parton's rapidity $y_\mathrm{init}$\footnote{In this work we impose the constraint $y_{\rm loss,2}\leq y_{\rm loss,4}\leq y_{\rm loss,6}$ in the Bayesian inference, which was introduced in Ref.~\cite{Jahan:2024wpj}.}, while the rapidity loss fluctuations are controlled by $\sigma_{y_\mathrm{loss}}$. For baryon charges, the 3D-Glauber model allows a probability $\lambda_B$ of assigning baryon charges to the string junction rather than at the string ends~\cite{Pihan:2024lxw}. Lastly, we estimate that the remnants of participant nucleons lose $\alpha_\mathrm{rem}$ fraction of rapidity compared to those of produced strings. The spatial positions of partons inside nucleons are sampled from a Gaussian with width $B_G$. The parameter $\alpha_\mathrm{shadowing}$ controls the number of strings produced from binary collisions. Strings produced in the 3D-Glauber model serve as energy-momentum source terms for the sequential hydrodynamic evolution. The spatial distributions of these source terms are specified by parameters $\sigma^\mathrm{string}_x$, $\sigma^\mathrm{string}_\eta$, and $\alpha_\mathrm{string\,tilt}$, which control the source term's transverse and longitudinal sizes, as well as the $x-\eta_s$ profile, respectively. Since these strings are decelerated along the longitudinal direction by $\tau_\mathrm{hydro} = 0.5$\,fm/$c$ before being deposited into hydrodynamic fields, we introduce the parameter $\alpha_\mathrm{preFlow}$ to take into account non-zero transverse flow developed during the string deceleration period~\cite{Zhao:2022ugy, Shen:2023awv}. In the hydrodynamic phase, the QGP specific shear viscosity is parameterized as a piecewise function in $\mu_B$ with the parameters $\eta_{0,2,4}$. The temperature dependence of the QGP's specific bulk viscosity is parameterized as an asymmetric Gaussian with parameters $\zeta_0, T_{\zeta_,0}, \sigma_{\zeta, \pm}$~\cite{Shen:2023awv}. Hydrodynamic fluid cells are converted into hadrons as soon as their local energy densities fall below the switching energy density $e_\mathrm{sw}$. 
For a more detailed description of the model and its parameters, we refer to Ref.~\cite{Jahan:2024wpj}, where we performed Bayesian inference using the RHIC BES data to obtain robust constraints on QGP transport properties.

The original training dataset for the Bayesian analysis was generated in Refs.~\cite{Shen:2023awv, jahan_2024_12807556, Jahan:2024wpj}.
We train the PCSK Gaussian Process (GP) emulator from the \texttt{surmise} package~\cite{surmise2023} as the fast surrogate model for this work. 
Compared to standard GP emulators, the PCSK GP takes into account the variation of statistical uncertainties across all training points~\cite{ankenman2010stochastic, Liyanage:2023nds}. 
A comparison of the different types of emulator performances was evaluated in Ref.~\cite{Roch:2024xhh}.

\begin{table}[t!]
    \caption{The experimental measurements for Au+Au collisions used in our Bayesian inference analysis. We include six species of identified hadrons ${\rm d}N/{\rm d}y$ and $\langle \pT \rangle$ from central to 60\% in collision centrality and the charged hadron $v_n\{2\}$ from 0 to 50\% centrality. We also include the normalized variances of charged hadron mean $\pT$, $\sigma_{\pT}^2/\langle \pT \rangle^2$ up to 60\% centrality and the $v_2\{\mathrm{SP}\}(\pT)$ data up to 60\% centrality.}
    \label{tab:training_data}
    \begin{tabular}{c|c|c}
        \hline\hline
        $\sqrt{s_{\rm NN}}$~[GeV] & STAR & PHOBOS \\
        \hline
        200 & $\mathrm{d}N/\mathrm{d}y (\pi^{\pm},K^{\pm},p,\bar{p})$~\cite{STAR:2008med} & $\mathrm{d}N_{\rm ch}/\mathrm{d}\eta$~\cite{PHOBOS:2005zhy} \\
            & $\langle \pT\rangle (\pi^{\pm},K^{\pm},p,\bar{p})$~\cite{STAR:2008med} & $v_2^{\rm ch}(\eta)$~\cite{PHOBOS:2006dbo} \\
            & $v_{2}^{\rm ch}\lbrace 2\rbrace$~\cite{STAR:2017idk}, $v_{3}^{\rm ch}\lbrace 2\rbrace$~\cite{STAR:2016vqt} \\
            & $\sigma_{\pT}^2/\langle \pT \rangle^2$~\cite{STAR:2019dow} & \\
            & $v_2\{\mathrm{SP}\}(\pT)$~\cite{STAR:2004jwm} & \\
        \hline
        19.6 & $\mathrm{d}N/\mathrm{d}y (\pi^{\pm},K^{\pm},p, \mathrm{net}\,p)$~\cite{STAR:2017sal} & $\mathrm{d}N_{\rm ch}/\mathrm{d}\eta$~\cite{PHOBOS:2005zhy} \\
             & $\langle \pT\rangle (\pi^{\pm},K^{\pm},p,\bar{p})$~\cite{STAR:2017sal} \\
             & $v_{2}^{\rm ch}\lbrace 2\rbrace$~\cite{STAR:2017idk}, $v_{3}^{\rm ch}\lbrace 2\rbrace$~\cite{STAR:2016vqt} \\
             & $\sigma_{\pT}^2/\langle \pT \rangle^2$~\cite{STAR:2019dow} & \\
             & $v_2\{\mathrm{SP}\}(\pT)$~\cite{STAR:2012och} & \\
        \hline
        7.7 & $\mathrm{d}N/\mathrm{d}y (\pi^{\pm},K^{\pm},p,\mathrm{net}\,p)$~\cite{STAR:2017sal} & \\
            & $\langle \pT\rangle (\pi^{\pm},K^{\pm},p,\bar{p})$~\cite{STAR:2017sal} & \\
            & $v_{2}^{\rm ch}\lbrace 2\rbrace$~\cite{STAR:2017idk}, $v_{3}^{\rm ch}\lbrace 2\rbrace$~\cite{STAR:2016vqt} \\
            & $\sigma_{\pT}^2/\langle \pT \rangle^2$~\cite{STAR:2019dow} & \\
            & $v_2\{\mathrm{SP}\}(\pT)$~\cite{STAR:2012och} & \\
        \hline\hline
    \end{tabular}
\end{table}

In this work, we will explore the potential optimization of the model-to-data calibration by introducing collision energy dependence on specific model parameters. 
We will also expand the scope of Bayesian analysis by introducing additional experimental measurements. 
In the following section, we will analyze how the posterior distributions change when more observables are added. 
The full list of experimental observables is summarized in Table~\ref{tab:training_data}. 
Compared to those used in Ref.~\cite{Jahan:2024wpj}, we include all six individual species of identified particle yields and their transverse momenta, normalized variances of charged hadron mean $\pT$, and $\pT$-differential elliptic flow $v_2\{\mathrm{SP}\}(\pT)$. 
For collisions at 19.6 and 7.7 GeV, we train emulators for protons and net proton yields, rather than those for protons and antiprotons, to reduce statistical fluctuations in the training datasets.

\section{Bayesian Model Selection}
\label{sec:BMS}

In this section, we optimize our theoretical model by introducing $\snn$-dependence on model parameters and utilize the Bayesian model selection method (Bayes' evidence) to determine whether the experimental measurements support such a model extension or not.

\subsection{Identify $\sqrt{s_{\rm NN}}$-dependent parameters}
\label{sec:energy_dependence_parameters}

One possible avenue to improve our previous analysis in Ref.~\cite{Jahan:2024wpj} without a substantial increase in computational cost is to introduce an explicit dependence on the collision energy $\snn$ for selected model parameters. 
This approach is motivated by previous findings of multimodal posterior distributions for some parameters in Ref.~\cite{Jahan:2024wpj}, which could be due to underlying energy-dependent behavior.

To systematically identify such dependencies, we perform three separate Bayesian analyses, each constrained to experimental data at a single collision energy: $\snn=200$, 19.6, and $7.7\;\mathrm{GeV}$, respectively.
Comparing the marginalized posterior distributions obtained from these independent analyses allows us to identify parameters exhibiting a $\snn$-dependent trend in a data-driven manner.

\begin{figure*}[t!]
    \includegraphics[width=\textwidth]{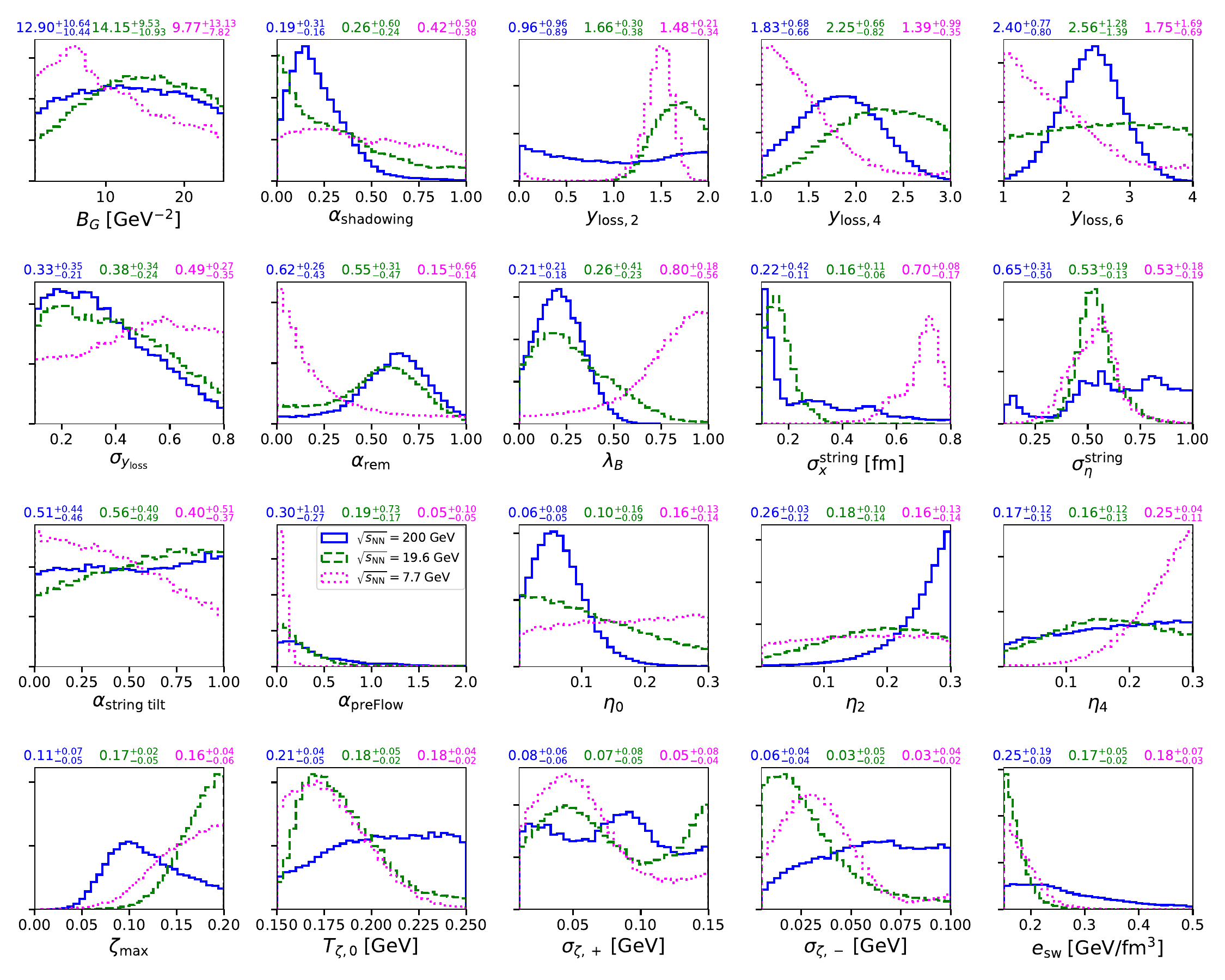}
    \caption{Marginal posterior distributions of the 20 model parameters obtained from separate Bayesian analyses at $\snn=200$, 19.6, and 7.7 GeV. The values above each panel indicate the median and $90\%$ credible intervals of the corresponding posterior distributions. The horizontal axes reflect the ranges of the uniform prior distributions used in the analyses.}
    \label{fig:posterior_comparison_sqrts}
\end{figure*}

Figure~\ref{fig:posterior_comparison_sqrts} shows the posterior distributions of all 20 model parameters across the three collision energies.
For several parameters, such as $y_{\rm loss,2}$, $y_{\rm loss,4}$, $y_{\rm loss,6}$, and $\eta_0$, $\eta_2$, $\eta_4$, an energy dependence is inherent by construction, as these parameters appear within implicitly $\snn$-dependent functional forms. 
For example, at $\snn=200\;\mathrm{GeV}$, the initial rapidity $y_{\rm init}$ is large, rendering $y_{\rm loss,2}$ effectively uninfluential; indeed, the posterior reflects this by remaining flat. 
Similarly, the parameter $\eta_4$, which controls the behavior of $\tilde{\eta}(\mu_B)$ at large $\mu_B$, is unconstrained at 200 GeV, consistent with the small net-baryon density at this energy.

In contrast, other parameters are not directly coupled to collision energy in their functional form but exhibit strong energy dependence in their inferred posteriors. 
For instance, the parameter $\alpha_{\rm rem}$, which controls the energy-momentum loss of collision remnants~\cite{Shen:2022oyg}, peaks near zero at 7.7 GeV, while the 19.6 and 200 GeV analyses favor higher values, $\alpha_{\rm rem} \approx 0.6$. 
This behavior is physically reasonable: even without substantial energy loss, the collision remnants overlap with the string source terms in the 3D-Glauber model at low collision energies.
Another example is $\lambda_B$, which controls baryon number fluctuations to string junctions~\cite{Pihan:2023dsb, Pihan:2024lxw}. 
Its energy dependence is in line with theoretical expectations, where the cross-sections for string junctions decrease with the collision energy~\cite {Kharzeev:1996sq}.

Similarly, the initial hotspot transverse size $\sigma_x$ is larger at 7.7 GeV and smaller at higher energies.
The more spread-out hotspots at low collision energy have less pressure gradients and develop less radial flow, which is constrained by the experimental mean $\pT$ measurements. Since the hotspot size represents the average spatial resolution in the collisions, its $\snn$ evolution indicates that the effective collision degree of freedom transits from partons to nucleons as the collision energy decreases. 

The particlization energy density $e_{\rm sw}$, which determines the transition from hydrodynamics to the hadronic transport model, also shows signs of energy dependence. 
At lower beam energies where $\mu_B$ is significant, the system is expected to depart from chemical equilibrium at a lower energy density, motivating a possible $\snn$-dependence of $e_{\rm sw}$.

In addition to this data-driven strategy, physics considerations can also guide the selection of energy-dependent parameters. 
For example, the parameter $\alpha_{\rm preFlow}$, which governs the initial transverse flow in the pre-equilibrium stage, could plausibly vary with $\snn$, as the duration and dynamics of the nuclear overlap change with beam energy.

In the next section, we systematically explore combinations of energy-dependent and independent parameters. 
By evaluating the model evidence using the Bayes factor, we identify the most favored configurations, thus providing a quantitative framework to determine which parameters benefit from explicit $\snn$-dependence.

\subsection{Bayesian model selection}
\label{sec:model_selection}
Having identified possible $\snn$-dependence in some model parameters in Sec.~\ref{sec:energy_dependence_parameters}, we now turn to Bayesian model selection to quantitatively compare competing model variants. In this case, our analysis includes data from all three Au+Au collision systems: 7.7 GeV, 19.6 GeV, and 200 GeV.

Considering two models $A$ and $B$, we can compute the posterior odds between the two models on the dataset $\mathbf{y}_{\rm exp}$ as~\cite{kass1995bayes}
\begin{align}
    \frac{\mathcal{P}(A \lvert \mathbf{y}_{\rm exp})}{\mathcal{P}(B \lvert \mathbf{y}_{\rm exp})} = \frac{\mathcal{P}(\mathbf{y}_{\rm exp}\lvert A)}{\mathcal{P}(\mathbf{y}_{\rm exp}\lvert B)} \frac{\mathcal{P}(A)}{\mathcal{P}(B)},
\end{align}
where the Bayes factor is defined as
\begin{align}
    \mathcal{B}_{A/B} = \frac{\mathcal{P}(\mathbf{y}_{\rm exp}\lvert A)}{\mathcal{P}(\mathbf{y}_{\rm exp}\lvert B)}
    \label{eq:B_evidence}
\end{align}
and the ratio $\mathcal{P}(A)/\mathcal{P}(B)$ is the prior odds between the two models. In our case, we do not introduce any bias on the models' priors, $\mathcal{P}(A) = \mathcal{P}(B)$, and leave the selection to the experimental data $\mathbf{y}_{\rm exp}$.

To compute the Bayes factor, we calculate the model evidence for a model $A$ as the likelihood marginalized over its parameter space:
\begin{equation}
    \mathcal{P}(\mathbf{y}_{\rm exp}\lvert A) = \int\mathrm{d}{\bs\theta}_{A}\;\mathcal{P}(\mathbf{y}_{\rm exp}\lvert {\bs\theta}_{A})\mathcal{P}({\bs\theta}_{A}).
    \label{eq:B_marginalized_L}
\end{equation}
We compute the model evidence using the \texttt{pocoMC} package~\cite{Karamanis:2022ksp,Karamanis:2022alw}, as employed in our Bayesian inference workflow~\cite{hendrik_roch_2025_15879411}.

As described in Sec.~\ref{sec:energy_dependence_parameters}, our analysis extends the dataset of Refs.~\cite{Shen:2023awv, Jahan:2024wpj} by including additional identified particle yields, larger rapidity coverage, transverse momentum fluctuation data, and $\pT$-differential anisotropic flow observables for all three systems. 
This results in a total of 858 experimental data points included in the inference.

Throughout this section, we refer to the baseline model with a 20-dimensional parameter space and no $\snn$ dependence (Tab.~\ref{tab:parameters}) as the ``no $\snn$'' setup. 
Variants with selected parameters allowed to depend on $\snn$ are treated as alternative models for comparison.

Table~\ref{tab:Bayes_factor} reports the Bayes factors for various such comparisons, where model $A$ is always the baseline model and model $B$ includes $\snn$ dependence for the parameters listed.
\begin{table}[t!]
    \caption{The natural logarithm of Bayes factors $\ln(\mathcal{B}_{A/B})$ comparing the baseline model $A$ (with no $\snn$ dependence) to model $B$ variants with energy-dependent parameters. Positive values favor the baseline, while negative values favor the extended model.}
    \label{tab:Bayes_factor}
    \centering
    \begin{tabular}{c|c|c}
        \hline\hline
        Model $A$ & Model $B$ & $\ln(\mathcal{B}_{A/B})$ \\
        \hline
        no $\sqrt{s_{\rm NN}}$ & $\alpha_{\rm shadowing}(\snn)$ & $0.88\pm 0.09$ \\
        no $\sqrt{s_{\rm NN}}$ & $\alpha_{\rm rem}(\snn)$ & $0.2\pm 0.3$ \\
        no $\sqrt{s_{\rm NN}}$ & $\lambda_B(\snn)$ & $0.2\pm 0.08$ \\
        no $\sqrt{s_{\rm NN}}$ & $\sigma_x(\snn)$ & $-2.00\pm 0.10$ \\
        no $\sqrt{s_{\rm NN}}$ & $\sigma_\eta(\snn)$ & $-1.6\pm 0.2$ \\
        no $\sqrt{s_{\rm NN}}$ & $\alpha_{\rm preFlow}(\snn)$ & $-0.2\pm 0.2$ \\
        no $\sqrt{s_{\rm NN}}$ & $e_{\rm sw}(\snn)$ & $0.57\pm 0.08$ \\
        no $\sqrt{s_{\rm NN}}$ & $\sigma_x(\snn), \sigma_\eta(\snn)$ & $-3.01\pm 0.08$ \\
        \hline\hline
    \end{tabular}
\end{table}
To interpret these results, we follow Jeffreys’ scale~\cite{Gordon:2007xm,Trotta2008}.\footnote{The interpretation of the Bayes factor follows Jeffreys' scale, where $\ln(\mathcal{B}_{A/B}) \gtrsim 5$ is considered strong evidence while $\ln(\mathcal{B}_{A/B}) \gtrsim 2.5$ is regarded as moderate evidence in favor of model~A.} 
Most model variants yield only weak or inconclusive evidence in favor of extending the parameterization to include $\snn$ dependence. 
However, models with energy dependence in the initial energy smearing parameters with only $\sigma_x$ or $\sigma_\eta$ show a strong preference, with $\ln(\mathcal{B}_{A/B})$ values around $-2$ indicating moderate evidence for the extended model.

This analysis highlights an important feature of Bayesian model selection: adding additional parameters does not automatically improve a model unless they are justified by the data. 
The Bayes factor provides a rigorous framework to penalize unnecessary complexity and reward genuine explanatory power.

Motivated by these findings, we also performed a Bayesian analysis where both $\sigma_x$ and $\sigma_\eta$ are modeled as functions of $\snn$. 
The resulting Bayes factor indicates a moderate preference for this more flexible model over the ``no $\snn$'' case.

The collision energy dependencies on initial hotspot sizes $\sigma_x$ and $\sigma_\eta$ were introduced in other phenomenological studies~\cite{Karpenko:2015xea, Schafer:2021csj}.

\subsection{$\mu_B$ dependence of shear and bulk viscosities}
\label{subsec:shear_bulk_viscosity}
Inspired by the analysis done in Ref.~\cite{JETSCAPE:2020mzn}, we can use the Bayes factor to investigate whether the experimental data can constrain the $\mu_B$-dependence of the QGP's specific viscosity. 
Since the baryon chemical potential $\mu_B$ varies with collision energy, a $\mu_B$-dependent specific bulk viscosity can effectively be translated to a dependence on $\snn$ within our model~\cite{Shen:2023awv}.

To explore this, we perform model comparisons with the base models $A$ as the setup favored in Sec.~\ref{sec:model_selection}, where the inclusion of $\sigma_x(\snn)$ and $\sigma_\eta(\snn)$ was supported by the data. 
The model $B$ extends the respective base model by either assuming a constant specific shear viscosity as a function of $\mu_B$, modeled by setting $\eta_0 = \eta_2 = \eta_4$, or allowing the maximum bulk viscosity $\zeta_{\rm max}$ to vary with $\snn$.

\begin{table}[t!]
    \caption{Bayes factors for model comparisons involving shear and bulk viscosities. Each row compares model $A$ to model $B$, where model $B$ includes an additional assumption: either a constant specific shear viscosity ($\eta_0=\eta_2=\eta_4$) or a $\snn$-dependent maximum bulk viscosity $\zeta_{\rm max}(\snn)$.}
    \label{tab:Bayes_factor_viscosity}
    \centering
    \begin{tabular}{c|c|c}
        \hline\hline
        Model $A$ & Model $B$ & $\ln(\mathcal{B}_{A/B})$ \\
        \hline
        $\eta_0, \eta_2, \eta_4$ & $\eta_0=\eta_2=\eta_4$ & $1.91\pm 0.07$ \\
        $\zeta_{\rm max}$ & $\zeta_{\rm max}(\snn)$ & $0.04\pm 0.06$ \\
        \hline\hline
    \end{tabular}
\end{table}

The results in Table~\ref{tab:Bayes_factor_viscosity} show that the RHIC BES data favors non-trivial $\mu_B$-dependence of the QGP specific shear viscosity as parameterized with $\eta_0$, $\eta_2$, $\eta_4$ in Refs.~\cite{Shen:2023awv, Jahan:2024wpj}. 
On the other hand, introducing a collision energy dependence of the bulk viscosity peak does not significantly improve the model-to-data calibration. 
Ref.~\cite{Shen:2023awv} showed a hint that the experimental data favored a non-monotonic $\snn$-dependence on $\zeta_\mathrm{max}$. 
This dependence can be compensated by the $\snn$-dependence on the hotspot transverse size $\sigma_x(\snn)$.

At this point, we have refined our optimized model to include the parameters listed in Table~\ref {tab:parameters} by adopting the $\snn$-dependent forms of $\sigma_x(\snn)$ and $\sigma_\eta(\snn)$ as favored by the Bayes factor analysis. 
This setup will be used in the following sections.

\section{Progressive constraints from different experimental data sets}
\label{sec:vary_exp_data}

In addition to modifying the model structure by introducing collision-energy-dependent parameters, we also investigate how the model posterior distribution is influenced by varying the amount of experimental data included in the Bayesian analysis.

\begin{table}[b!]
    \caption{The list of posterior distributions calibrated with different experimental datasets.}
    \label{tab:posteriorList}
    \centering
    \begin{tabular}{c|c}
        \hline\hline
        Posterior Name & Experimental dataset \\ \hline
        Posterior 1 & Ref.~\cite{Jahan:2024wpj} \\ \hline
        Posterior 2 & Tab.~\ref{tab:training_data} without $v_2\{\mathrm{SP}\}(\pT)$ \\ \hline
        Posterior 3 & Tab.~\ref{tab:training_data} \\
        \hline\hline
    \end{tabular}
\end{table}

As listed in Table~\ref{tab:posteriorList}, we have three different setups that progressively include additional experimental observables in the Bayesian calibration. 
We begin with the posterior distribution obtained in our previous work~\cite{Jahan:2024wpj}. 
In the second Bayesian calibration, we expand our scope to include STAR measurements of six identified particle species' yields and their mean transverse momenta at RHIC BES energies~\cite{STAR:2017sal}. 
We also include the normalized variance of charged hadron transverse momenta measurements from STAR~\cite{STAR:2019dow}. 
Finally, we further include the $\pT$-differential elliptic flow measurements (see Table~\ref{tab:training_data})~\cite{STAR:2004jwm, STAR:2012och}.

\begin{figure*}[t!]
    \includegraphics[width=\textwidth]{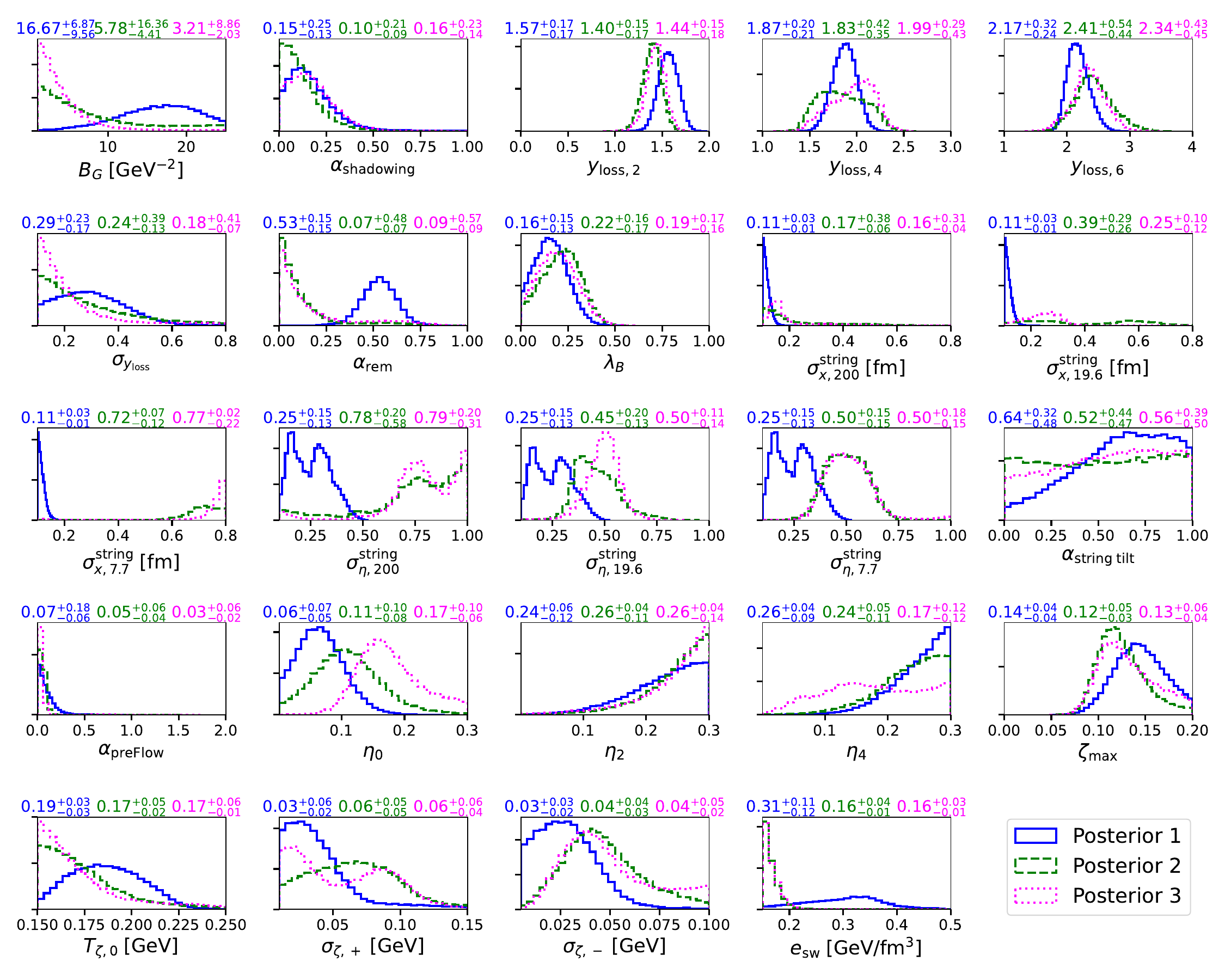}
    \caption{Marginal posterior distributions of the 24 model parameters obtained from separate Bayesian analyses with three different experimental datasets. The values above each panel indicate the median and $90\%$ credible intervals of the corresponding posterior distributions. The horizontal axes reflect the ranges of the uniform prior distributions used in the analyses.}
    \label{fig:posterior_comparison_data}
\end{figure*}

Figure~\ref{fig:posterior_comparison_data} compares the 1D marginal distributions for individual model parameters from the three posterior distributions described in Table~\ref{tab:posteriorList}. 
The original Posterior 1 from Ref.~\cite{Jahan:2024wpj} does not include any $\snn$-dependent parameters. 
Therefore, we repeat the posterior distributions of $\sigma_x$ and $\sigma_\eta$ to compare with $\sigma_x(\snn)$ and $\sigma_\eta(\snn)$ in the other two posterior distributions, respectively. 
Figure~\ref{fig:posterior_comparison_data} shows that the distributions of a few parameters have significant changes between Posterior 1 and 2. 
We verified that introducing the $\snn$-dependent initial-state parameters $\sigma_x(\snn)$ and $\sigma_\eta(\snn)$, while keeping the same experimental data in the inference, does not have significant effects on the model's posterior distribution. 
The shifts in distributions present in Fig.~\ref{fig:posterior_comparison_data} are from the additional experimental data constraints.

From Posterior 1 to 2, the extra experimental constraints come from the yields of additional identified particles.
They directly lead to a lower switching energy density $\esw$ for particlization from hydrodynamics to the hadronic transport description. 
Because all the model parameters are correlated in the Bayesian inference, this constraint also caused other model parameters to change. 
The preferred values of the nucleon width parameter $B_G$ shift to smaller values when comparing Posterior 1 to the other two distributions, changing the granularity of transverse energy profiles. 
Along the longitudinal direction, we find that the optimal initial-state energy-loss fraction for the participant nucleon remnants $\alpha_\mathrm{rem}$ reduces from 0.5 in Posterior 1 to nearly zero in the other two distributions. 
This shift is correlated to the preference for large values of $\sigma_\eta(\snn)$ in Posterior 2 and 3. 
A larger value of $\sigma_\eta$ smears the initial energy density profile along the longitudinal direction, compensating for the effects from the energy loss of the participant nucleon remnants.

Allowing the hotspot size to vary at different collision energies, we find that the peak value of $\sigma_x$ increases with decreasing $\snn$ in Posterior 2 and 3. 
This trend is consistent with previous phenomenological studies~\cite{Karpenko:2015xea, Schafer:2021csj}. 
Moreover, the parameter $\sigma_x$ controls the sizes of initial pressure gradients in the transverse plane. 
Its $\snn$-dependence reduces the significance of non-monotonic bulk viscosity in the $\mu_B$~\cite{Shen:2023awv} as seen in Table~\ref{tab:Bayes_factor_viscosity}.
Looking at the $\snn$ dependence of $\sigma_\eta$, we find the preferred value of $\sigma_\eta$ decreases with the collision energy, in line with the previous phenomenological studies~\cite{Karpenko:2015xea, Schafer:2021csj, Shen:2020jwv}.

To further quantify improvements in data-to-model comparisons, we perform numerical simulations with model parameters sampled from the posterior distributions. 
We estimate the theoretical uncertainties in the observables with the variance of our results from using different parameter sets~\cite{Jahan:2024wpj}. 
However, because performing event-by-event simulations with many different parameter sets is computationally expensive, we can only conduct high-statistical simulations at a limited number of parameter sets (five in this work). 
To obtain reasonable estimations of the theoretical uncertainties in model predictions with the available computational resources, we adopt a cluster sampling approach~\cite{henderson1982cluster, roesch1993adaptive}. 
We perform clustering on the posterior distribution and sample the model parameter sets from different clusters to ensure they are sufficiently distinct in the parameter space. 
The details of the cluster sampling procedure are discussed in the Appendix~\ref{app:posterior_clustering}. 

\begin{figure}[h!]
    \centering
    \includegraphics[width=\linewidth]{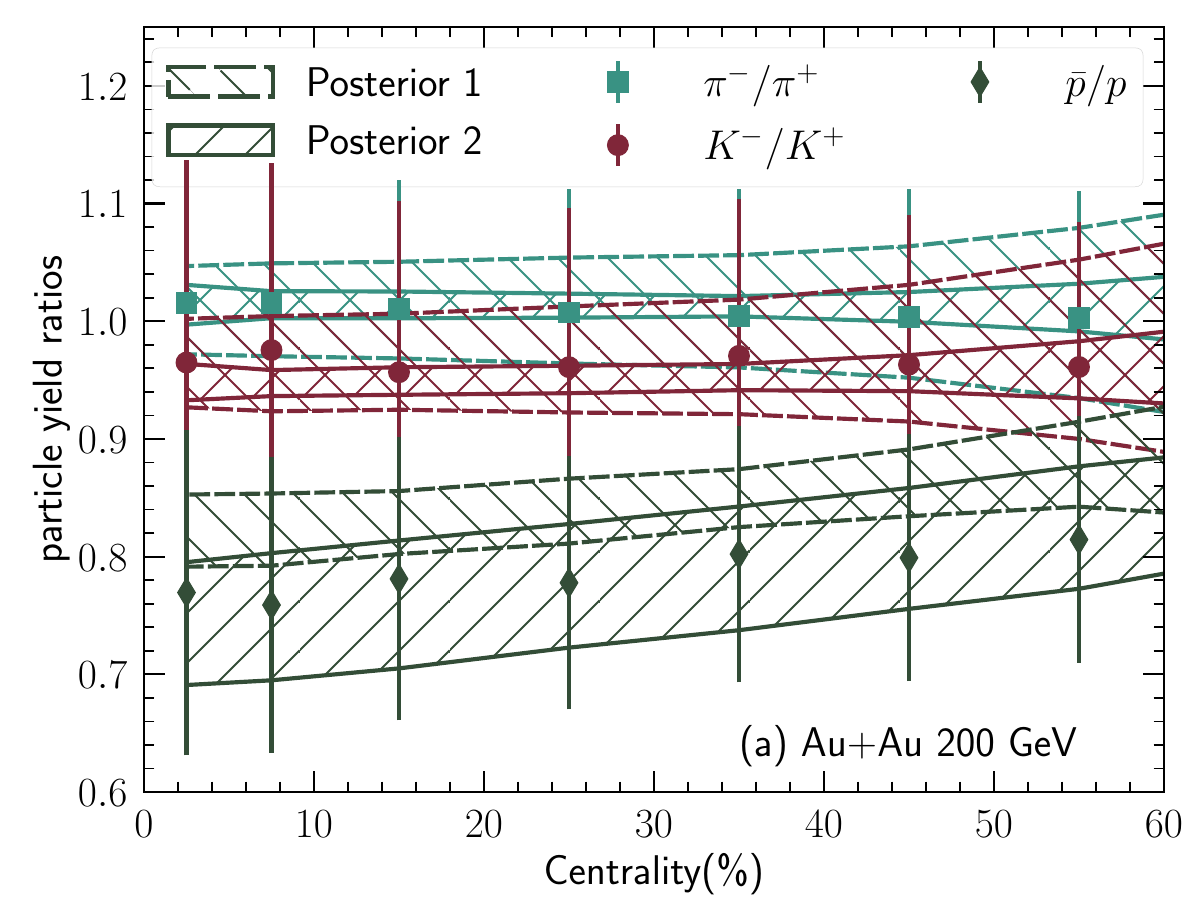}
    \includegraphics[width=\linewidth]{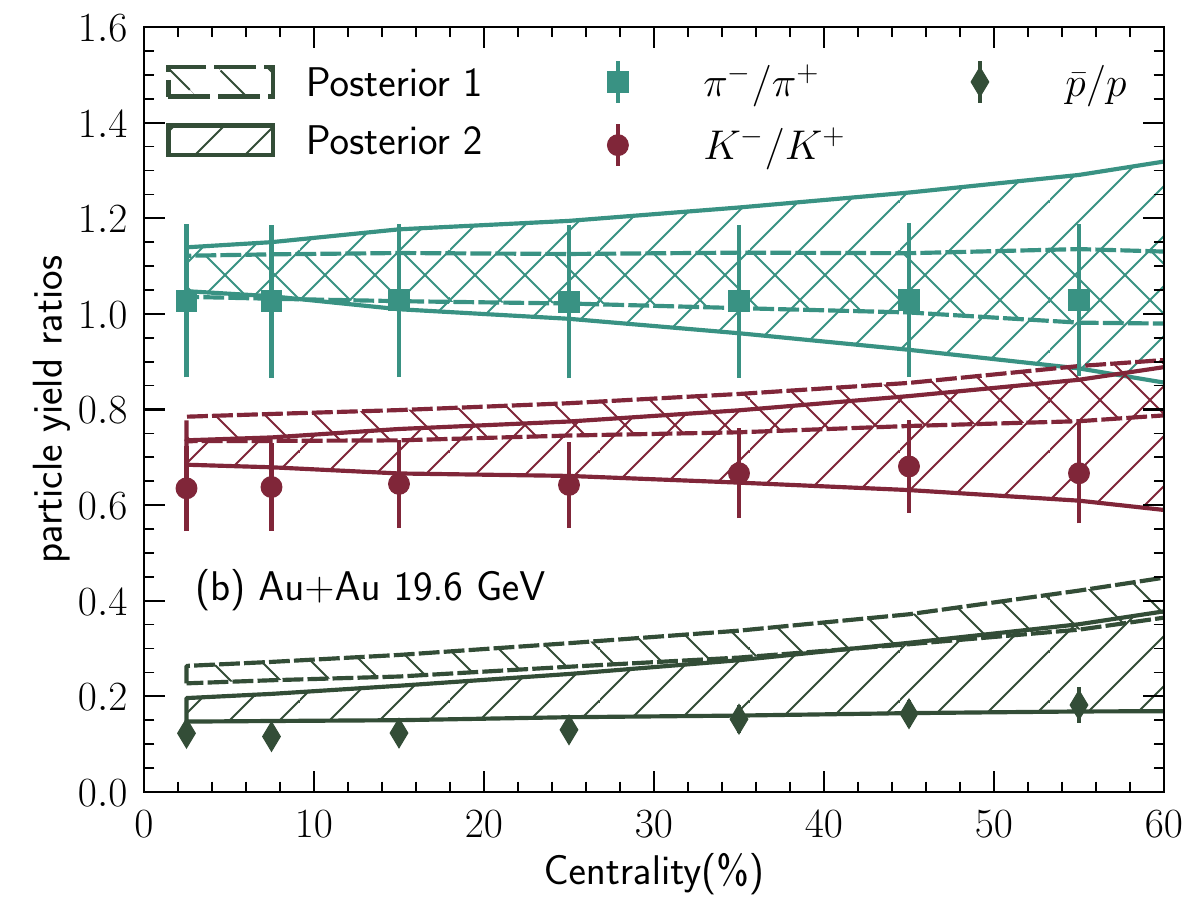}
    \includegraphics[width=\linewidth]{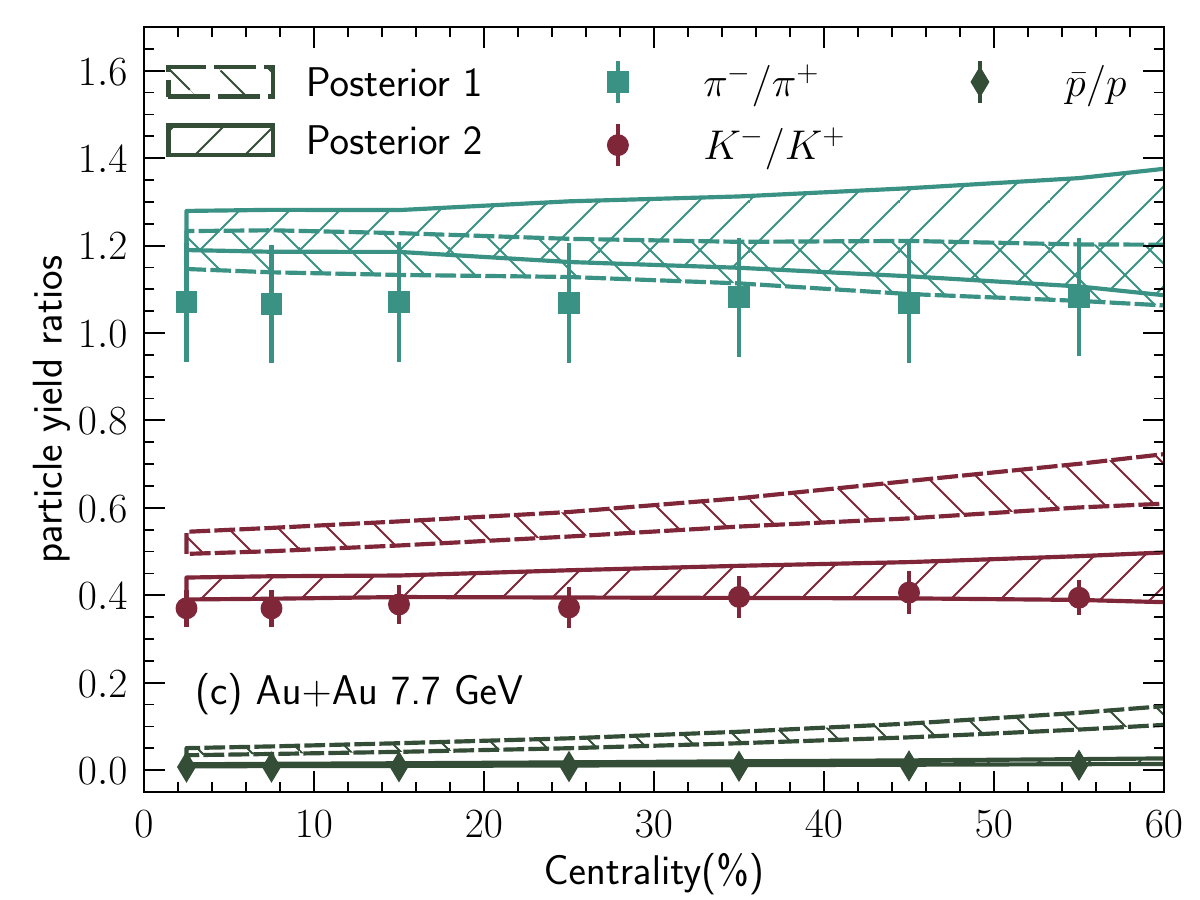}
    \caption{Centrality dependence of identified particle yields from simulations with Posterior distributions 1 and 2 compared with the STAR measurements for Au+Au collisions at 200 (panel (a)), 19.6 (panel (b)), and 7.7 (panel (c)) GeV. The shaded bands represent systematic uncertainty in the theoretical results.}
    \label{fig:posterior_comparison_piddNRatios}
\end{figure}

Figure~\ref{fig:posterior_comparison_piddNRatios} presents the comparisons of anti-particle to particle yield ratios with the STAR measurements in the RHIC BES program. 
The shaded bands represent the systematic uncertainties ($\pm$ two standard deviations) of the model simulations using five parameter sets sampled from the posterior distributions.
We find that a high value of the switching energy density $\esw \approx 0.31$\,GeV/fm$^3$ in Posterior 1 results in overestimation of the $K^-/K^+$ and $\bar{p}/p$ ratios for heavy-ion collisions with $\snn \le 20$\,GeV. Once we include the yields of $\pi^-$, $K^-$, and net protons (for better emulator accuracy) in the Bayesian inference, the model-to-data comparison is significantly improved by using a low value of the switching energy density $\esw \approx 0.16$\,GeV/fm$^3$. This result is consistent with previous studies of hadronic chemistry in Pb+Pb collisions at the top SPS energy~\cite{Monnai:2019hkn, Monnai:2020pcw}.

\begin{figure}[h!]
    \centering
    \includegraphics[width=\linewidth]{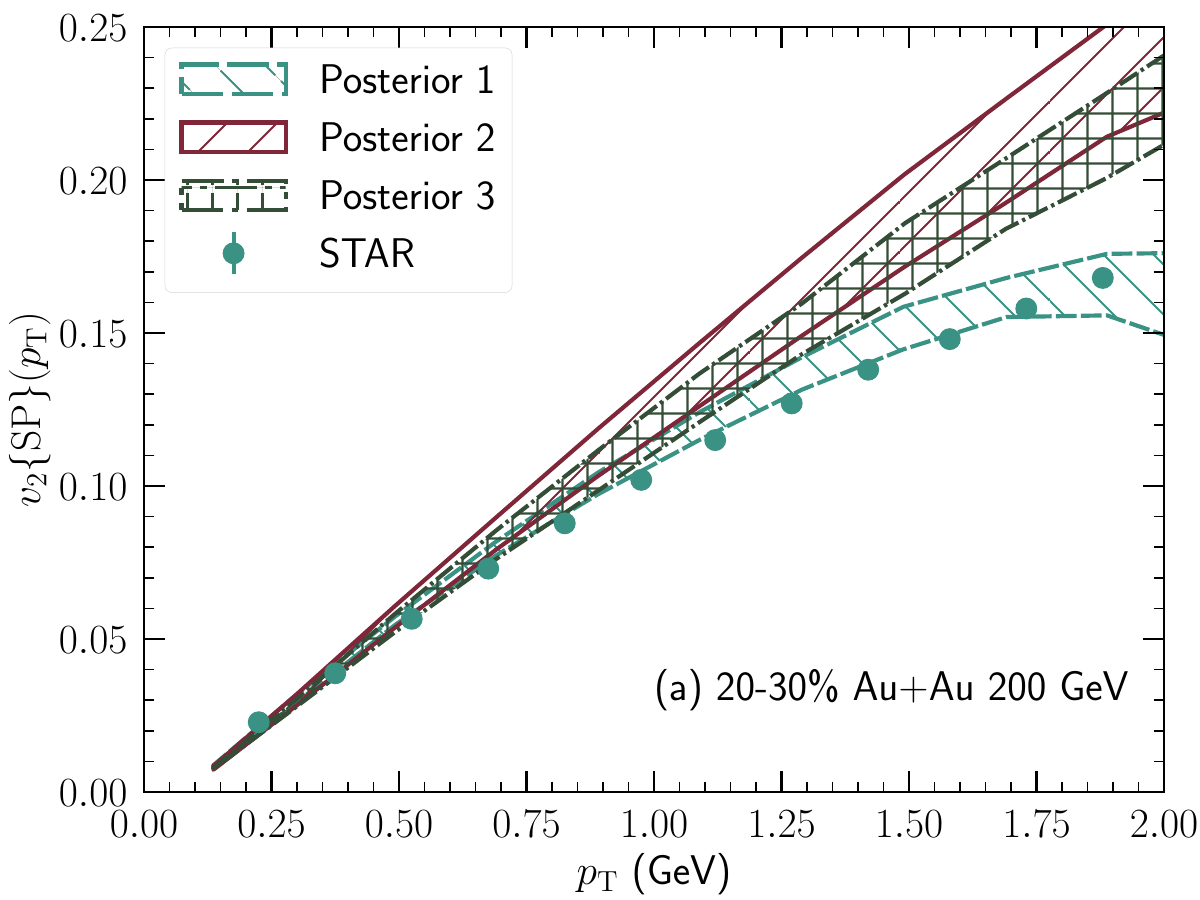}
    \includegraphics[width=\linewidth]{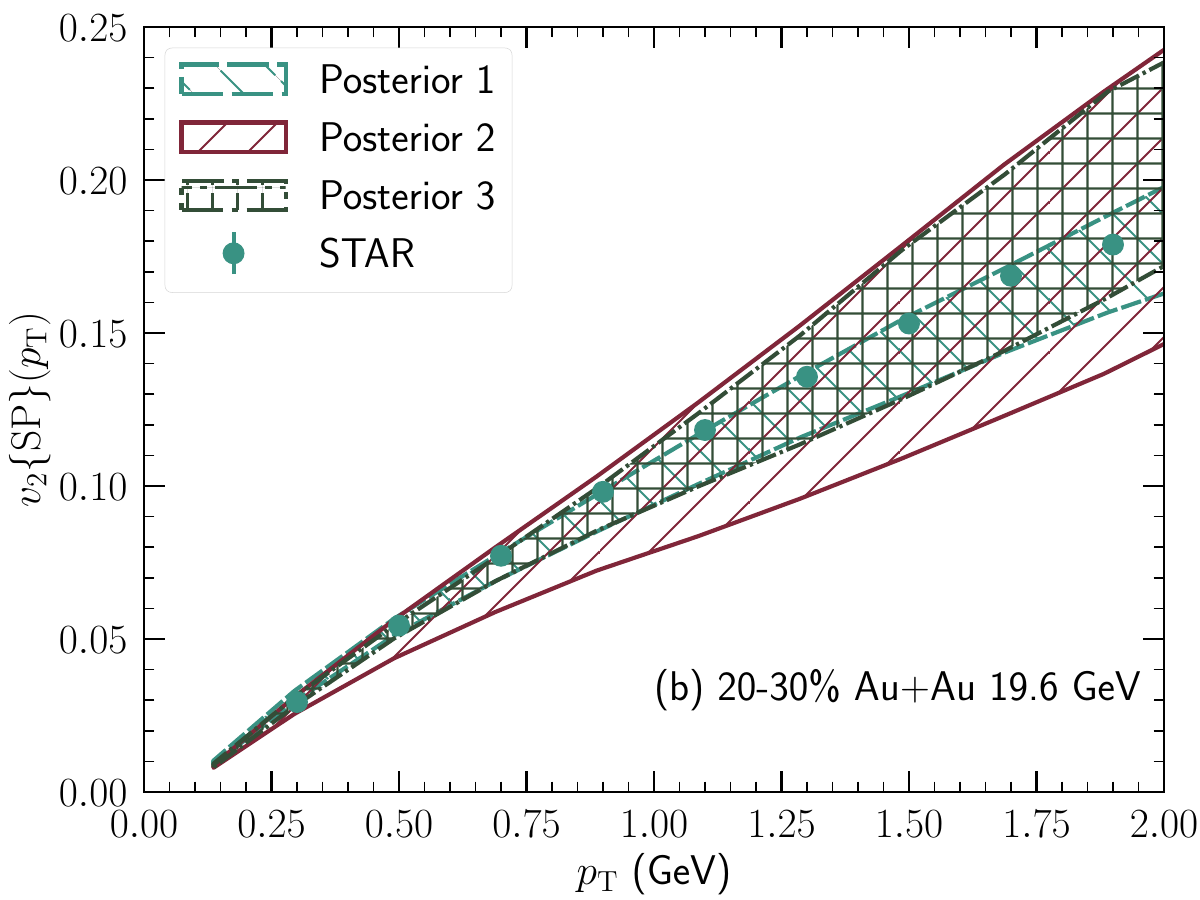}
    \includegraphics[width=\linewidth]{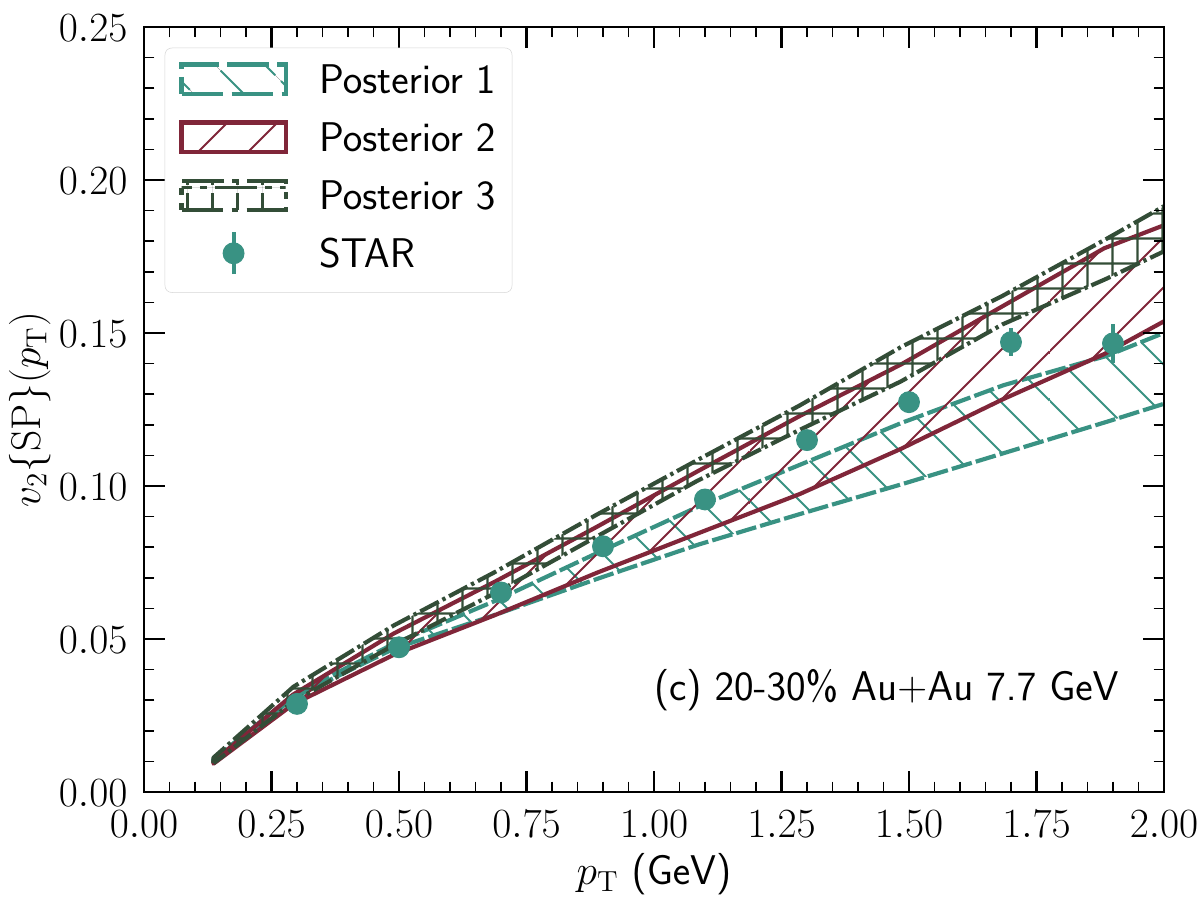}
    \caption{Charged hadron $v_2(\pT)$ from simulations with three Posterior distributions compared with the STAR measurements for 20-30\% Au+Au collisions at 200 (panel (a)), 19.6 (panel (b)), and 7.7 (panel (c)) GeV. The shaded bands represent systematic uncertainty in the theoretical results.}
    \label{fig:posterior_comparison_v2pT}
\end{figure}

However, the shift of the posterior distribution of $\esw$ towards the low values from Posterior 1 to 2 also triggers significant changes in a few other model parameters in the Bayesian inference. 
We find that the preferred value of the specific shear viscosity $\eta_0$ at $\mu_B = 0$ is shifted to a larger value. 
This change can be understood as the lower switching energy density leads to a longer lifetime for hydrodynamic evolution of the system, which develops more radial and anisotropic flow in the hadronic phase. 
Therefore, a larger specific shear viscosity is required to reproduce a similar amount of anisotropic flow in the STAR measurements.

In the meantime, we find that hybrid simulations with $\esw \approx 0.16$\,GeV/fm$^3$ lead to an overestimation of the $\pT$-differential elliptic flow at 200 GeV, as shown in Figure~\ref{fig:posterior_comparison_v2pT}. 
This is because the out-of-equilibrium corrections ($\delta f$) at the particlization are smaller at lower values of $\esw$~\cite{Shen:2010uy}. 
If we further include the $\pT$-differential elliptic flow measurements in the Bayesian calibration, the posterior distribution for the specific shear viscosity parameter $\eta_0$ further shifts to larger values in Posterior 3.
However, the model simulations sampled from Posterior 3 only slightly improve the description of $v_2(\pT)$ compared to the results from Posterior 2, indicating tensions with other experimental data.

In Appendix~\ref{Sec:deltaf}, we quantify the effects of out-of-equilibrium correlations ($\delta f$) on experimental observables in the RHIC BES energies.
A comprehensive comparison between the posterior model simulations and all calibration data in Tab.~\ref{tab:training_data} is presented in Appendix~\ref{app:BayesianCalibration}.

\section{Model predictions with uncertainty propagation}
\label{sec:prediction}

In this section, we provide some model predictions for upcoming measurements requested by the STAR Collaboration.
We also make comparisons with the existing measurements for rapidity-dependent anisotropic flow $v_n(\eta)$. 
The theoretical uncertainties on these observables are estimated using the cluster sampling method from the posterior distributions discussed in Sec.~\ref{sec:vary_exp_data}.
When making predictions for observables at collision energies which are not used in the Bayesian calibration, we keep all the $\snn$-independent model parameters fixed and determine the values of $\sigma_x(\snn)$ and $\sigma_\eta(\snn)$ by linear interpolation with $\ln(\snn)$ from neighbouring values in the parameter sets.

\subsection{Longitudinal flow decorrelation}

The longitudinal flow decorrelation is a key experimental observable that constrains initial-state fluctuations and the (3+1)D dynamical evolution~\cite{CMS:2015xmx, ATLAS:2017rij}.
For symmetric collision systems, the anisotropic flow rapidity correlation function is defined as
\begin{align}
    r_n(\eta) = \frac{\langle Q_n(-\eta) Q_n^*(\eta^B_\mathrm{ref}) + Q_n(\eta) Q_n^*(\eta^A_\mathrm{ref}) \rangle}{\langle Q_n(\eta) Q_n^*(\eta^B_\mathrm{ref}) + Q_n(-\eta) Q_n^*(\eta^A_\mathrm{ref}) \rangle},
    \label{eq:rn}
\end{align}
where the complex anisotropic flow vector is defined as
\begin{align}
    Q_n \equiv \sum_j w_j \exp[-i n \phi_j]
    \label{eq:Qn}
\end{align}
with the index $j$ summing over all charged hadrons within the given kinematic cuts. 
The particle weight $w_j$ is usually set to unity for most of the analyses. 
To compare with the upcoming STAR measurements, we use charged hadrons with $\pT \in [0.4, 4]$\,GeV and $|\eta| < 1.5$ for the anisotropic flow vector $Q_n(\eta)$. 
The reference flow vectors $Q_n(\eta^{A,B}_\mathrm{ref})$ were computed with charged hadrons in the same $\pT$ range, but $-5.1 < \eta^A_\mathrm{ref} < -3.1$ and $3.1 < \eta^B_\mathrm{ref} < 5.1$ for $r_2(\eta)$ and $-5.1 < \eta^A_\mathrm{ref} < -2.1$ and $2.1 < \eta^B_\mathrm{ref} < 5.1$ for $r_3(\eta)$.

\begin{figure}[t!]
    \centering
    \includegraphics[width=\linewidth]{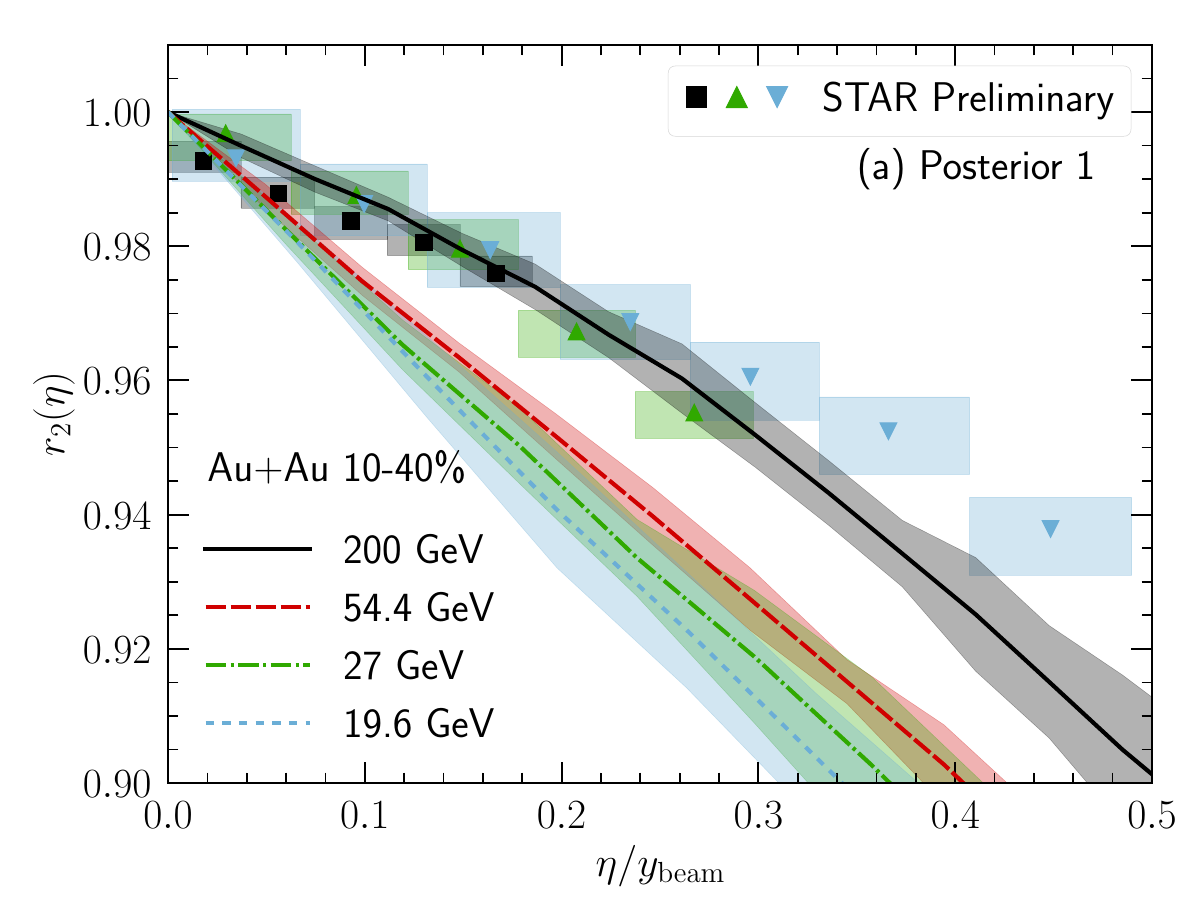}
    \includegraphics[width=\linewidth]{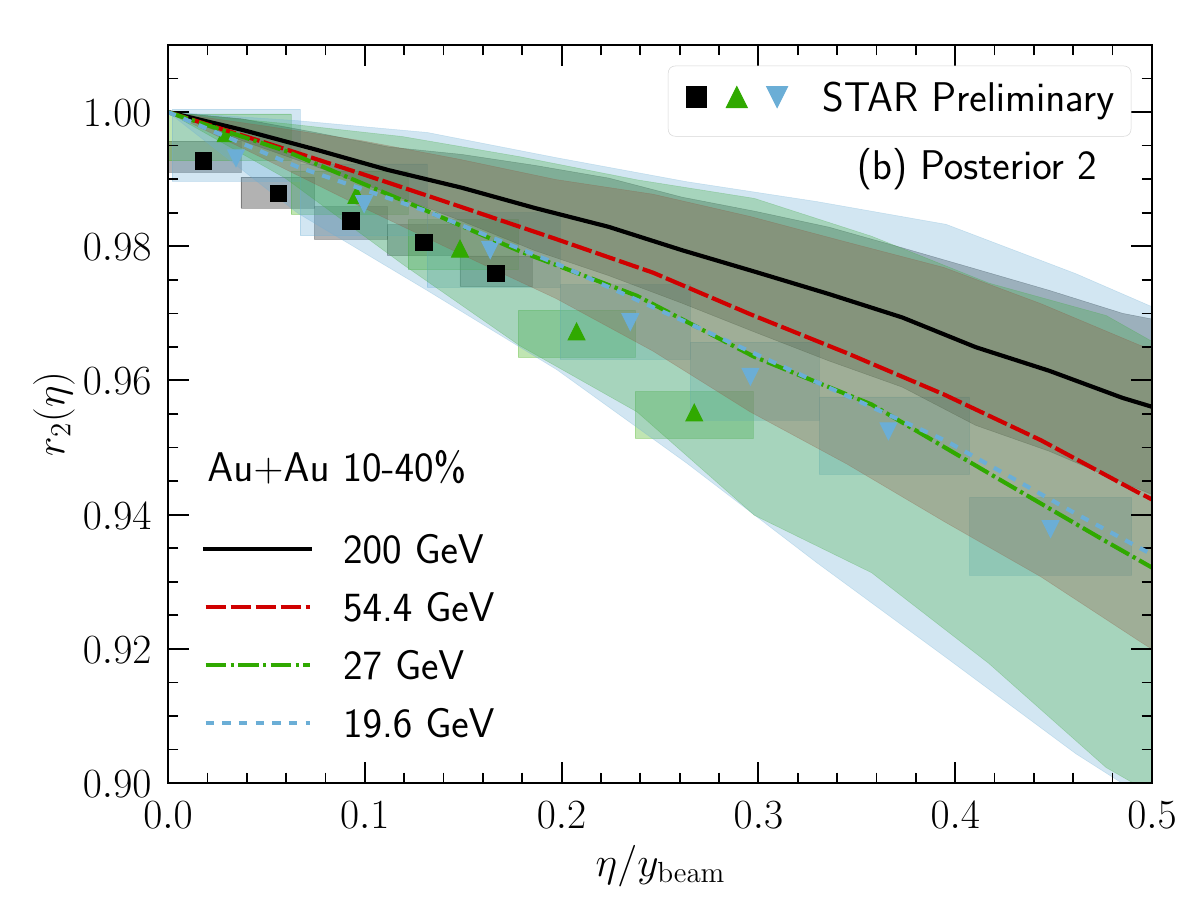}
    \caption{The longitudinal decorrelation for elliptic flow $r_2(\eta)$ in 10-40\% Au+Au collisions at RHIC BES program scaled by the corresponding beam rapidity. The preliminary STAR measurements are compared with model results from Posterior 1 (Panel (a)) and Posterior 2 (Panel (b)). The shaded bands represent systematic uncertainty in the theoretical results. The STAR preliminary data are extracted from the STAR talks~\cite{STAR_rn_QMtalks}.}
    \label{fig:PostPred_r2}
\end{figure}

Figures~\ref{fig:PostPred_r2}a and b show our model results for the elliptic flow rapidity correlation $r_2(\eta)$ scaled by the collision beam rapidity for 10-40\% Au+Au collisions with $\snn =19.6 - 200$\,GeV. 
We do not show the results from Posterior 3 because they are consistent with those obtained using Posterior 2. 
The difference between the results from Posterior 1 and 2 is mainly driven by the switching energy density $\esw$ in our model. 
Lower values of $\esw$ from Posterior 2 allow for a longer fireball lifetime in the hydrodynamic phase, resulting in more alignment of the elliptic flow vectors $Q_2(\eta)$ at different rapidities and bringing $r_2(\eta)$ closer to unity. 
With $\esw \approx 0.16$\,GeV/fm$^3$ in Posterior 2, our calculations of $r_2(\eta)$ from different collision energies roughly scale with the variable $\eta/\ybeam$ within the theoretical uncertainties, which is indicated by the STAR preliminary measurements~\cite{STAR_rn_QMtalks}. 
This scaling behavior breaks down for higher values of $\esw$, as seen in results using Posterior 1.

\begin{figure}[t!]
    \centering
    \includegraphics[width=\linewidth]{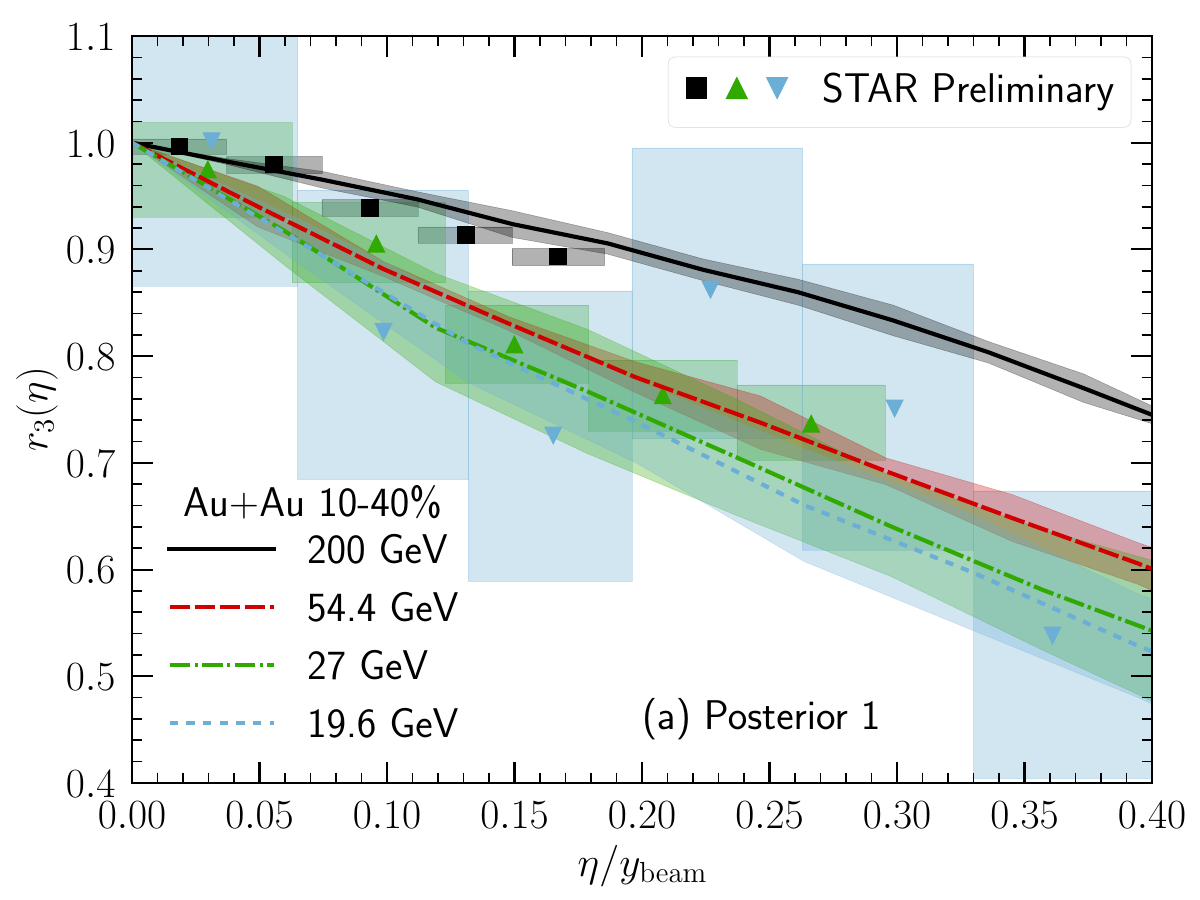}
    \includegraphics[width=\linewidth]{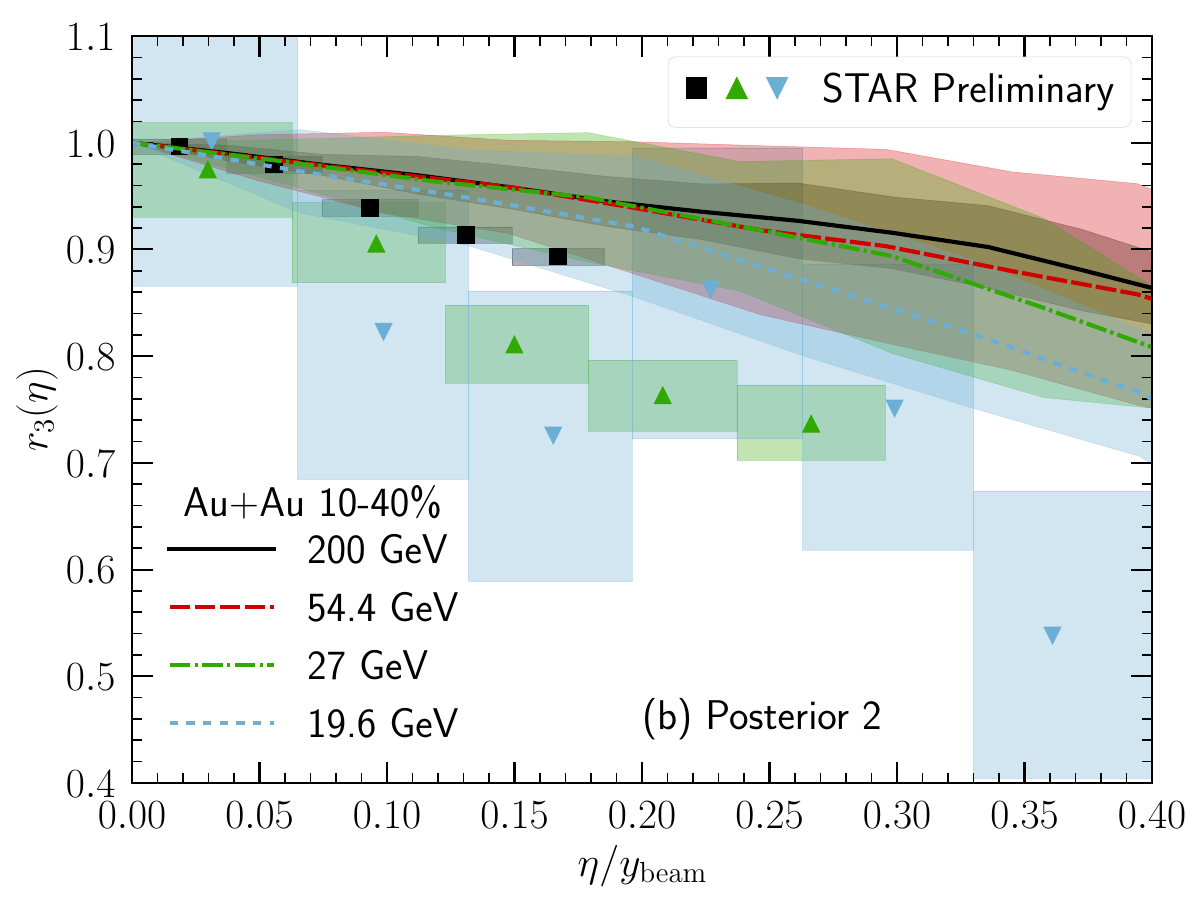}
    \caption{Similar to Fig.~\ref{fig:PostPred_r2} but for $r_3(\eta)$ for 10-40\% Au+Au collisions compared with the preliminary STAR measurements~\cite{STAR_rn_QMtalks}.}
    \label{fig:PostPred_r3}
\end{figure}

Figures~\ref{fig:PostPred_r3} further present our results for the triangular flow decorrelation $r_3(\eta)$, scaled by the collision beam rapidity. 
Our model results show qualitatively the same dependences as those in $r_2(\eta)$ between the two posterior distributions. 
Compared to the preliminary STAR measurements, the $r_3(\eta)$ from Posterior 2 underestimates the amount of longitudinal flow decorrelation. 
The preliminary STAR measurements for $r_3(\eta)$ indicate a small scaling breakdown with $\ybeam$, which agrees better with our model results from Posterior 1. 

Overall, the strong sensitivity of our model posteriors to $r_n(\eta)$ observables makes the upcoming precision measurements a valuable observable to impose additional constraints on our model parameters in future Bayesian analyses.

\subsection{Collision energy dependence of $v_2(\eta)$}

The STAR Collaboration will measure new rapidity-dependent anisotropic flow with its Event Plane Detector (EPD) in the Beam Energy Scan program~\cite{STAR_vneta_talk}. 
The event-plane $v_n\{\mathrm{EP}\}(\eta)$ is defined as follows with the anisotropic flow vectors,
\begin{align}
    v_n\{\mathrm{EP}\}(\eta) = \frac{\langle V_n(\eta) \exp(- i n \Psi_n) \rangle}{\mathrm{Res}(\Psi_n)},
\end{align}
where the normalized anisotropic flow vector $V_n(\eta) \equiv Q_n(\eta)/Q_0(\eta)$ and $\Psi_n \equiv (1/n)\mathrm{Arg}(Q_n)$ is the $n$-th order event plane angle, and $\mathrm{Res}(\Psi_n)$ is the event-plane resolution. 
To suppress non-flow contributions, the $v_n\{\mathrm{EP}\}(\eta)$ in STAR's Time Projection Chamber (TPC) coverage ($|\eta| < 1.5$) uses the event plane determined by the EPD detector ($2.1 < |\eta| < 5.1$). 
Its event-plane resolution is computed with the 3sub-event method as follows,
\begin{align}
    \mathrm{Res}(\Psi_n^\mathrm{EPD}) =& \bigg[\langle \cos[n(\Psi_n^\mathrm{TPC, E} - \Psi_n^\mathrm{EPD})] \rangle \nonumber \\
    & \quad \times \langle \cos[n(\Psi_n^\mathrm{TPC, W} - \Psi_n^\mathrm{EPD})] \rangle \nonumber \\
    & \quad / \langle \cos[n(\Psi_n^\mathrm{TPC, E} - \Psi_n^\mathrm{TPC, W})] \rangle \bigg]^{1/2},
\end{align}
where the anisotropic flow vectors in the two TPC sub-events are computed with charged hadrons with $\pT \in [0.1, 4]$\,GeV and rapidity ranges $-1.5 < \eta < -0.5$ (TPC, W) and $0.5 < \eta < 1.5$ (TPC, E).
Similarly, the $v_n\{\mathrm{EP}\}(\eta)$ in STAR's EPD coverage ($2.1 < |\eta| < 5.1$) uses the event plane determined by the TPC detector. 
Its event plane resolution is computed as
\begin{align}
    \mathrm{Res}(\Psi_n^\mathrm{TPC}) =& \bigg[\langle \cos[n(\Psi_n^\mathrm{EPD, E} - \Psi_n^\mathrm{TPC})] \rangle \nonumber \\
    & \quad \times \langle \cos[n(\Psi_n^\mathrm{EPD, W} - \Psi_n^\mathrm{TPC})] \rangle \nonumber \\
    & \quad / \langle \cos[n(\Psi_n^\mathrm{EPD, E} - \Psi_n^\mathrm{EPD, W})] \rangle \bigg]^{1/2},
\end{align}
where the two sub-events in the EPD are defined with rapidity coverages $-5.1 < \eta < -2.8$ (EPD, W) and $2.8 < \eta < 5.1$ (EPD, E) for all particles without any $\pT$ cuts, and the reference flow vector in TPC is computed using charged hadrons with $\pT \in [0.1, 4]$\,GeV and rapidity $0.5 < |\eta| < 1.5$.

Although we follow the experimental description to compute these observables, we note that precise comparisons can not be made because the anisotropic flow measurements of the event plane method potentially contain uncontrolled bias from flow fluctuations~\cite{Luzum:2012da}.

\begin{figure}[h!]
    \centering
    \includegraphics[width=\linewidth]{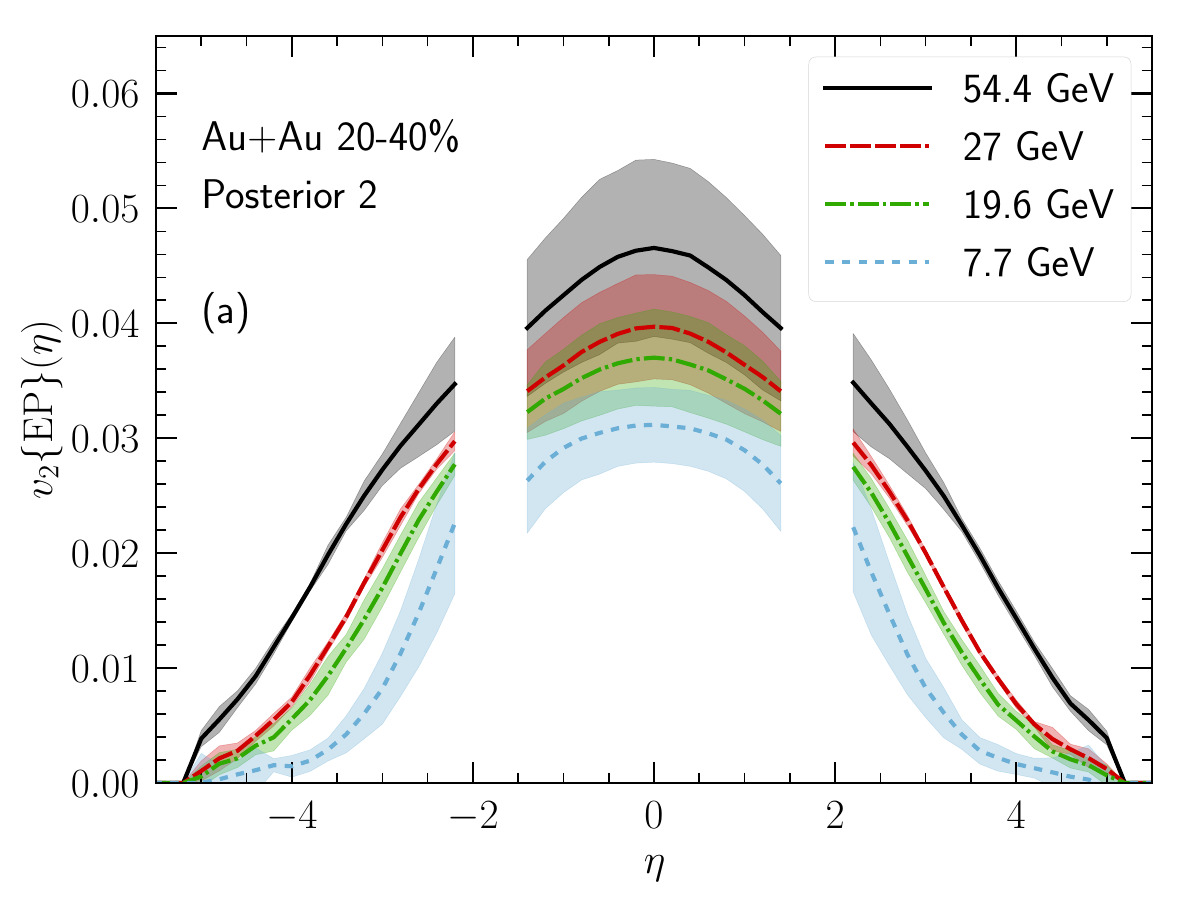}
    \includegraphics[width=\linewidth]{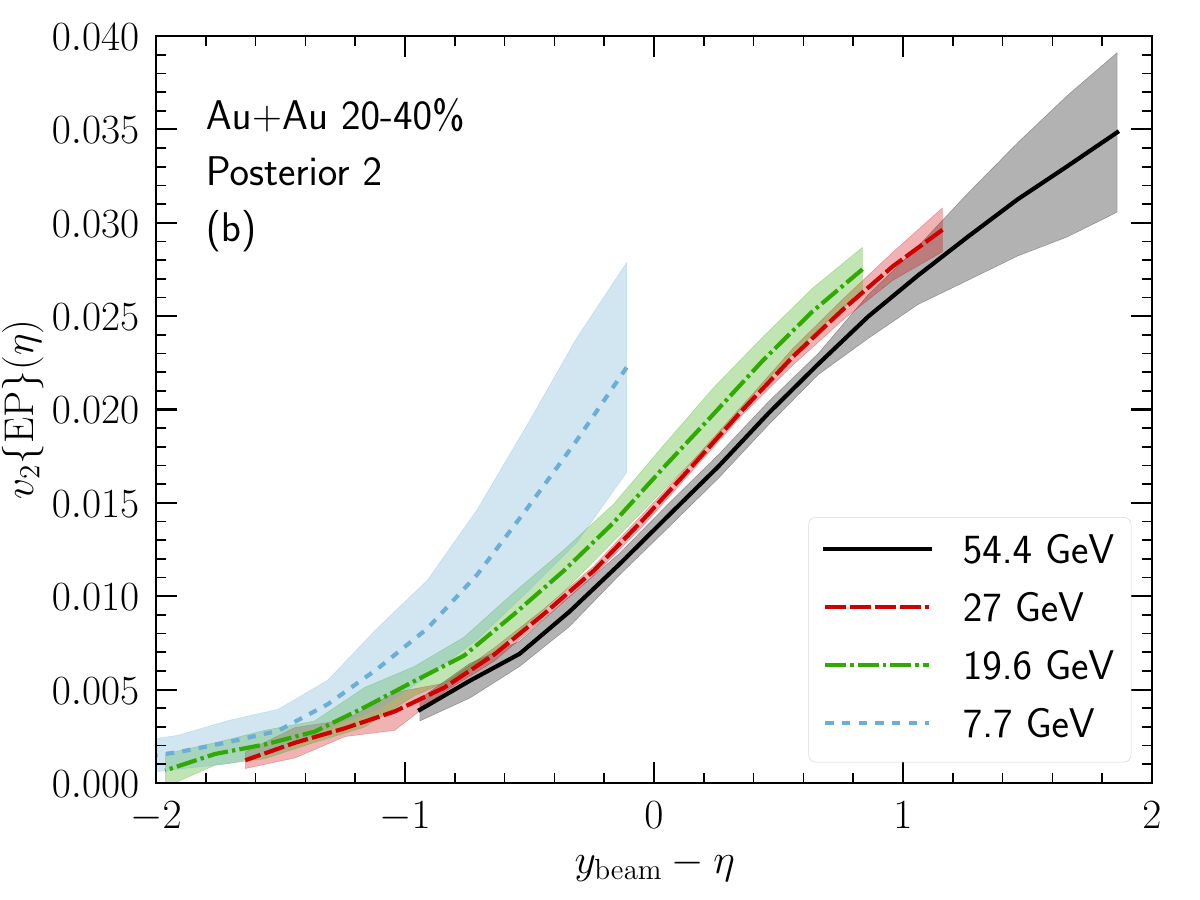}
    \caption{The event-plane pseudorapidity dependent elliptic flow of charged hadrons for 20-40\% Au+Au collisions in the RHIC BES program. Panel (a) displays the results in pseudorapidity, while panel (b) shifts the results by their corresponding beam rapidities in the fragmentation region. The shaded bands represent systematic uncertainty in the theoretical results.}
    \label{fig:PostPred_vneta}
\end{figure}

Figure~\ref{fig:PostPred_vneta}a shows our model predictions for the event-plane elliptic flow of charged hadrons as a function of pseudorapidity in 20-40\% Au+Au collisions, spanning an energy range from 54.4 to 7.7 GeV.
The predictions for $v_2\{\mathrm{EP}\}(\eta)$ are made in the TPC ($|\eta| < 1.5$) and EPD ($2 < |\eta| < 5.1$) rapidity ranges, which can be compared with the upcoming STAR measurements. 
Our results from the three posterior distributions are consistent with each other within their uncertainties. 
Therefore, we will only present the results from Posterior 2 here.

In Fig.~\ref{fig:PostPred_vneta}b, we shift the $v_2\{\mathrm{EP}\}(\eta)$ from different collision energies by their corresponding beam rapidity. 
We observe an approximate scaling in the fragmentation region for $v_2\{\mathrm{EP}\}(\eta)$ in $-1 < \ybeam -\eta < 1$ for Au+Au collisions between 54.4 and 19.6 GeV. 
The elliptic flow at $\snn = 7.7$ GeV does not follow this scaling. 
When $\eta$ is close to the beam rapidity, we note that there are non-trivial contributions from the collision spectators, which we do not include in our model simulations.

The PHOBOS Collaboration measured the pseudo-rapidity dependence of elliptic flow with the event-plane method~\cite{PHOBOS:2004vcu}. 
To compare with PHOBOS measurements, we computed $v_2\{\mathrm{EP}\}(\eta)$ using charged hadrons with transverse momenta $\pT \in [0.15, 2]$\,GeV. 
The event plane angle and its resolution are computed using charged hadrons with the same $\pT$ cut in the rapidity range $2.05 < |\eta| < 3.2$ measured by the octagonal multiplicity detector~\cite{PHOBOS:2004vcu}. 
We have included the PHOBOS 0-40\% measurements~\cite{PHOBOS:2006dbo} in the Bayesian calibration (see Table~\ref{tab:training_data}). 

\begin{figure}[t!]
    \centering
    \includegraphics[width=\linewidth]{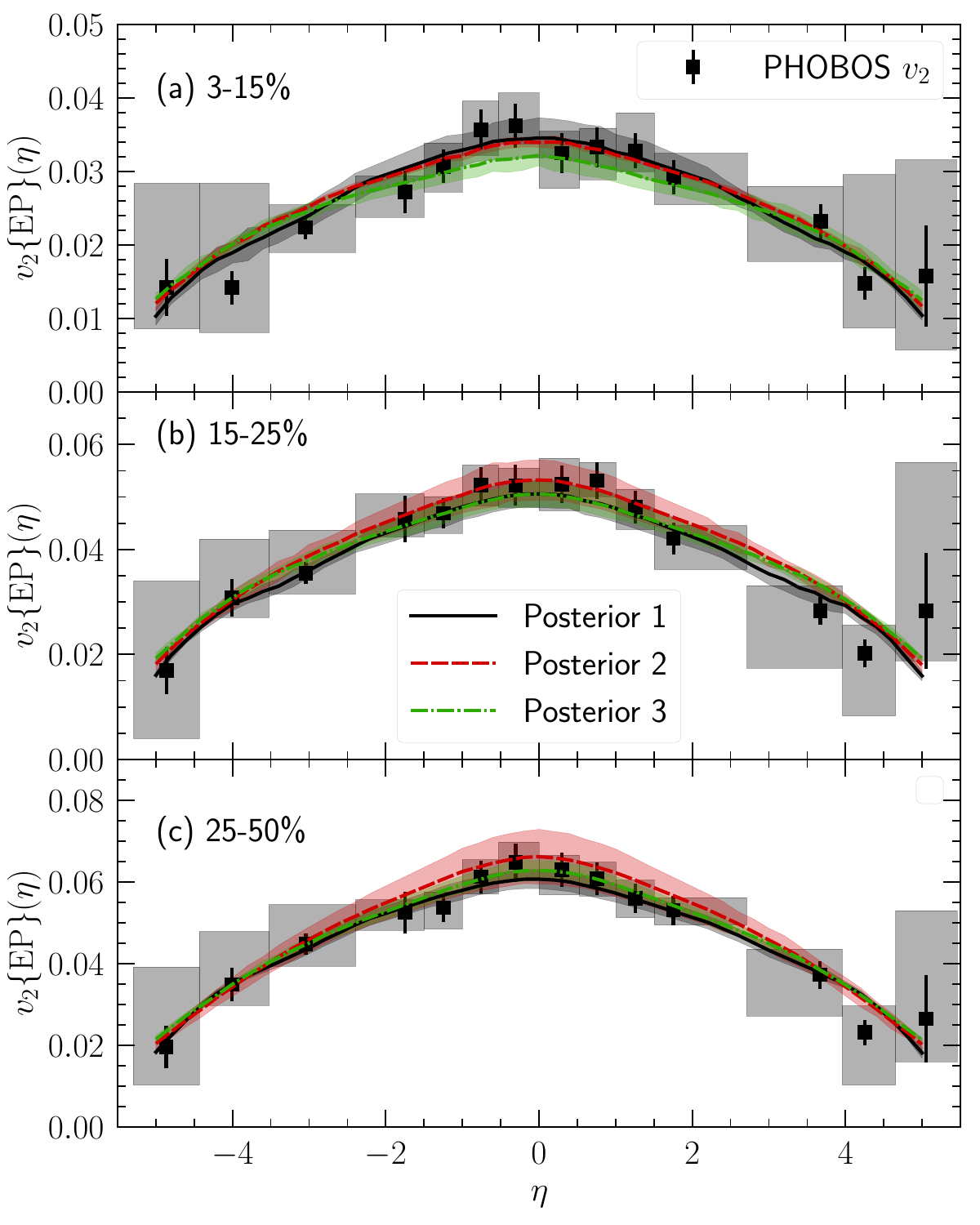}
    \caption{The pseudorapidity dependent charged hadron elliptic flow $v_2(\eta)$ in Au+Au collisions at 200 GeV compared with the PHOBOS measurements for 3-15\% (a), 15-25\% (b), and 25-50\% (c) centrality bins~\cite{PHOBOS:2004vcu}. The shaded bands represent systematic uncertainty in the theoretical results.}
    \label{fig:v2etaPHOBOS}
\end{figure}

Figure~\ref{fig:v2etaPHOBOS} shows the comparison with the PHOBOS measurements in three centrality bins of Au+Au collisions at $\snn = 200$\,GeV. 
Our model results, based on the three posterior distributions, are consistent with each other. 
They reproduce the pseudorapidity dependence of the elliptic flow. 
The triangle-shaped $\eta$ dependence in $v_2\{\mathrm{EP}\}$ was the main observable that results in the increasing QGP's specific shear viscosity with $\mu_B$~\cite{Shen:2023pgb}.

\begin{figure}[h!]
    \centering
    \includegraphics[width=\linewidth]{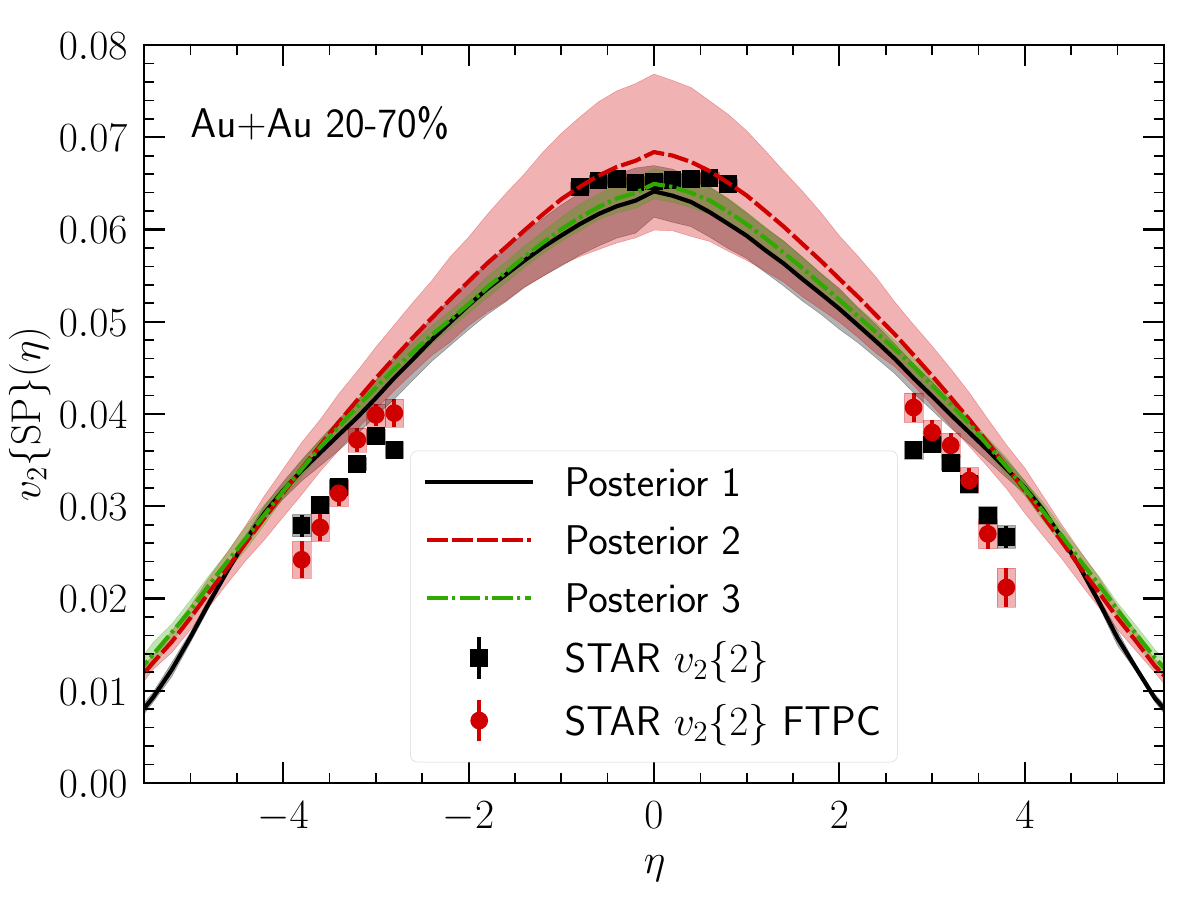}
    \caption{The pseudorapidity dependent charged hadron elliptic flow $v_n\{\mathrm{SP}\}(\eta)$ with the scalar-product method in 20-70\% Au+Au collisions at 200 GeV compared with the STAR measurements~\cite{STAR:2004jwm}. The shaded bands represent systematic uncertainty in the theoretical results.}
    \label{fig:v2etaSTAR200}
\end{figure}

The STAR Collaboration also measured the pseudorapidity dependent elliptic flow at 200 GeV~\cite{STAR:2004jwm}. 
To compare with this STAR measurement, we compute the rapidity-dependent anisotropic flow with the scalar-product (SP) method $v_n\{\mathrm{SP}\}(\eta)$ as follows,
\begin{align}
    v_n\{\mathrm{SP}\}(\eta) = \left\{\begin{array}{lr}
    \frac{\langle V_n(\eta) Q_n^*(\eta_\mathrm{ref}^A) \rangle}{\sqrt{\langle Q_n(\eta_\mathrm{ref}^A) Q_n^*(\eta_\mathrm{ref}^B)\rangle}} & \quad\mbox{for } \eta > 0 \\
    \frac{\langle V_n(\eta) Q_n^*(\eta_\mathrm{ref}^B) \rangle}{\sqrt{\langle Q_n(\eta_\mathrm{ref}^A) Q_n^*(\eta_\mathrm{ref}^B)\rangle}} & \quad\mbox{for } \eta < 0
    \end{array} \right. ,
    \label{eq:vnSP}
\end{align}
where the reference anisotropic flow $Q_n$ vectors are computed from two sub-events ($A$ and $B$) with $-1 < \eta_\mathrm{ref}^A < -0.5$ and $0.5 < \eta_\mathrm{ref}^B < 1$ for the STAR TPC detector or $-4.2 < \eta_\mathrm{ref}^A < -2.4$ and $2.4 < \eta_\mathrm{ref}^B < 4.2$ for the STAR Forward Time Projection Chamber (FTPC) detectors~\cite{STAR:2004jwm}. 
In this case, the reference anisotropic flow vectors $Q_n(\eta_\mathrm{ref})$ are computed with the weight $w_j = p_{\mathrm{T}, j}$ of individual hadrons in Eq.~\eqref{eq:Qn} according to Ref.~\cite{STAR:2004jwm}.
All the anisotropic flow vectors are computed using charged hadrons with transverse momenta $\pT \in [0.15, 2]$\,GeV.
We find that the $v_n^\mathrm{sub}(\eta)$ calculated using TPC or FTPC reference flow vectors yields very close results from our model. 
Therefore, in Fig.~\ref{fig:v2etaSTAR200}, we only show the $v_2\{\mathrm{SP}\}(\eta)$ with the FTPC reference flow vectors, which have large rapidity gaps.

Figure~\ref{fig:v2etaSTAR200} shows our model comparison with the STAR measurements in 20-70\% Au+Au collisions at 200 GeV.\footnote{Based on the description in Ref.~\cite{STAR:2004jwm}, it is not clear whether the measurements were made directly in such a wide centrality bin or averaged over measurements from multiple narrow centrality bins.} 
The STAR measurements are significantly more precise than the PHOBOS measurements. 
In this case, we observe that the model $v_2\{\mathrm{SP}\}(\eta)$ decreases more slowly in the forward/backward pseudorapidity regions than in the data. 
This discrepancy suggests that introducing a temperature-dependent $\eta/s$ with increasing values at lower temperatures would improve the model-to-data comparison~\cite{Shen:2020jwv}. 

\begin{figure}[h!]
    \centering
    \includegraphics[width=\linewidth]{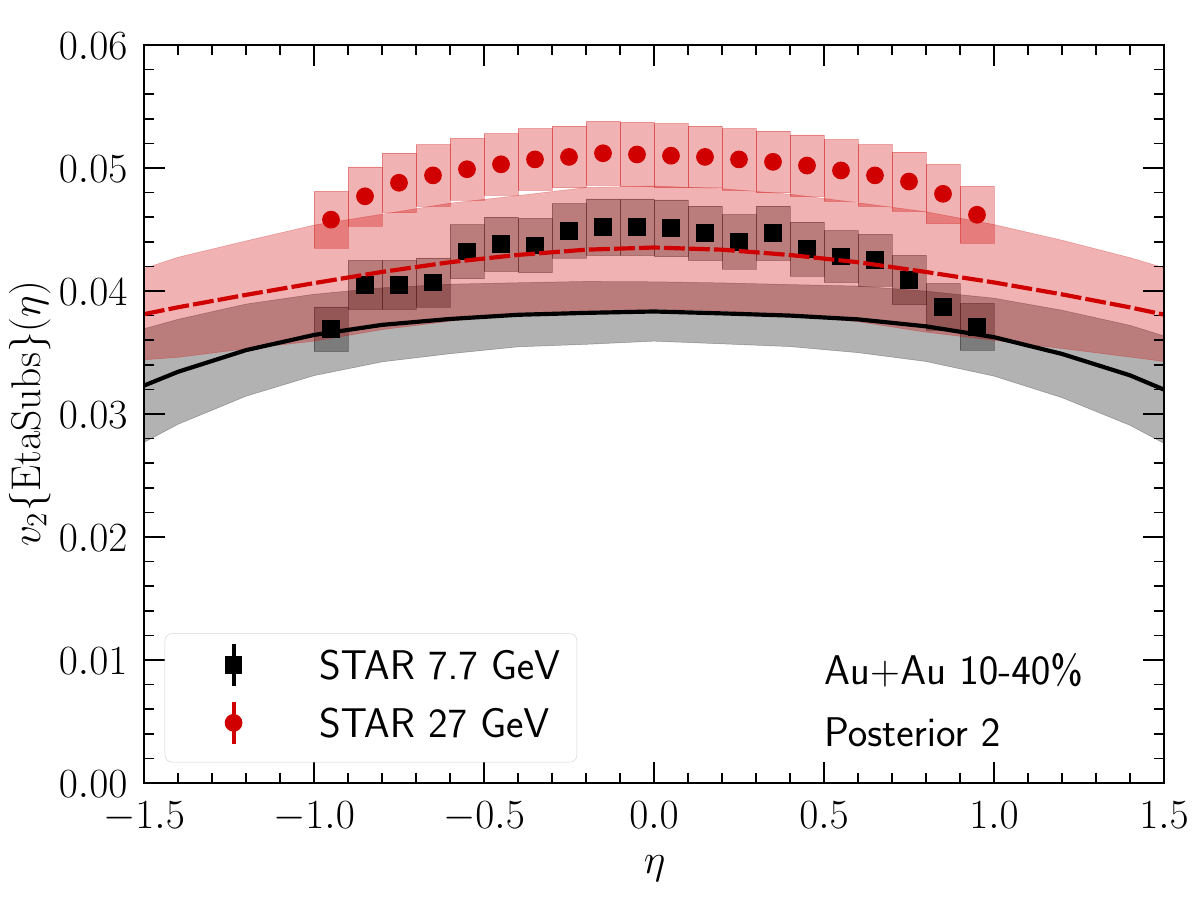}
    \caption{The pseudorapidity dependent charged hadron elliptic flow $v_2\{\mathrm{EtaSubs}\}(\eta)$ in 10-40\% Au+Au collisions at 27 and 7.7 GeV compared with the STAR measurements~\cite{STAR:2012och}. The shaded bands represent systematic uncertainty in the theoretical results.}
    \label{fig:v2etaSTARBES}
\end{figure}

At the BES energies, the STAR Collaboration measured the pseudorapidity-dependent elliptic flow with the $\eta$ sub-event method $v_2\{\mathrm{EtaSubs}\}(\eta)$ within a limited rapidity window $| \eta | < 1$~\cite{STAR:2012och}.
The $\eta$ sub-event method elliptic flow is defined similarly to Eq.~\eqref{eq:vnSP}, but replacing the reference flow vectors $Q_n(\eta^{A, B}_\mathrm{ref})$ by only their unit vector $\exp(-i n \Psi_{n, \mathrm{ref}}^{A, B}) \equiv Q_n(\eta^{A, B}_\mathrm{ref})/|Q_n(\eta^{A, B}_\mathrm{ref})|$.
This $\eta$ sub-event method is a variation of the event-plane method, and it also contains uncontrolled bias from flow fluctuations~\cite{Luzum:2012da}.
In this measurement, the two sub-events ($A$ and $B$) were chosen as $-1 < \eta^A_\mathrm{ref} < -0.075$ and $0.075 < \eta^B_\mathrm{ref} < 1$
from the STAR TPC detector~\cite{STAR:2012och}.\footnote{The pseudorapidity gap between the two sub events is only $|\Delta \eta| \ge 0.15$ in this analysis. It is not clear whether all the non-flow contributions have been removed from the elliptic flow measurements.} 
All the anisotropic flow vectors are computed with charged hadrons with transverse momenta $\pT \in [0.2, 2]$\,GeV.

Figure~\ref{fig:v2etaSTARBES} shows the comparison of our model results with the STAR measurements in 10-40\% Au+Au collisions at $\snn =27$ and 7.7 GeV. 
Unlike the comparisons at 200 GeV, we underestimate the STAR $v_2\{\mathrm{EtaSubs}\}(\eta)$ by 10\% in the TPC region. 
It indicates that the current constrained specific shear viscosity at finite $\mu_B$ (related to the model parameters $\eta_2$ and $\eta_4$ in Table~\ref{tab:parameters}) is too large. 
However, such specific shear viscosity was required by the forward $v_2(\eta)$ measurements at 200 GeV. 
This tension can be released by introducing a temperature-dependent $\eta/s$ in future Bayesian calibration.

Overall, the precise measurements of pseudorapidity-dependent anisotropic flow $v_n(\eta)$ are valuable observables to constrain the temperature and $\mu_B$ dependence of the QGP specific shear viscosity~\cite{Denicol:2015nhu, Shen:2020jwv, Gotz:2025wnv}.

\subsection{Anisotropic flow in small systems}

In this subsection, we make predictions for the anisotropic flow coefficients in O+O and d+Au collision systems. 
The upcoming measurements at RHIC can provide valuable insights into whether hydrodynamic responses to initial geometry are the origin of anisotropic flow in small systems~\cite{Giacalone:2025vxa}. 

We compute the two-particle anisotropic flow coefficients $v_n\{2\}$ using charged hadrons with $\pT \in [0.2, 2]$\,GeV, which can be measured in the STAR TPC detector.
With event-by-event anisotropic flow vectors in a given centrality bin, we compute $v_n\{2\}$ with two subevents $A$ and $B$,
\begin{align}
    v_n\{2\} = \sqrt{\frac{\langle Q_n(\eta_\mathrm{ref}^A) Q^*_n(\eta_\mathrm{ref}^B) \rangle}{\langle Q_0(\eta_\mathrm{ref}^A) Q^*_0(\eta_\mathrm{ref}^B) \rangle}},
    \label{eq:star_2_event_rapidity}
\end{align}
where the denominator is the average number of particle pairs.
Experimentally, rapidity gaps are imposed to suppress non-flow contributions in the two-particle correlation. 
Because the anisotropic flow vectors depend on rapidity, especially for asymmetric d+Au collisions, we impose the same rapidity gap as the experiments with the two sub-events. 
The flow vectors $Q_n$ are computed using charged hadrons with $-1.5 < \eta_\mathrm{ref}^A < -0.5$ and $0.5 < \eta_\mathrm{ref}^B < 1.5$, which impose a minimum pseudorapidity gap of $|\Delta \eta| \ge 1$.

\begin{figure}[h!]
    \centering
    \includegraphics[width=\linewidth]{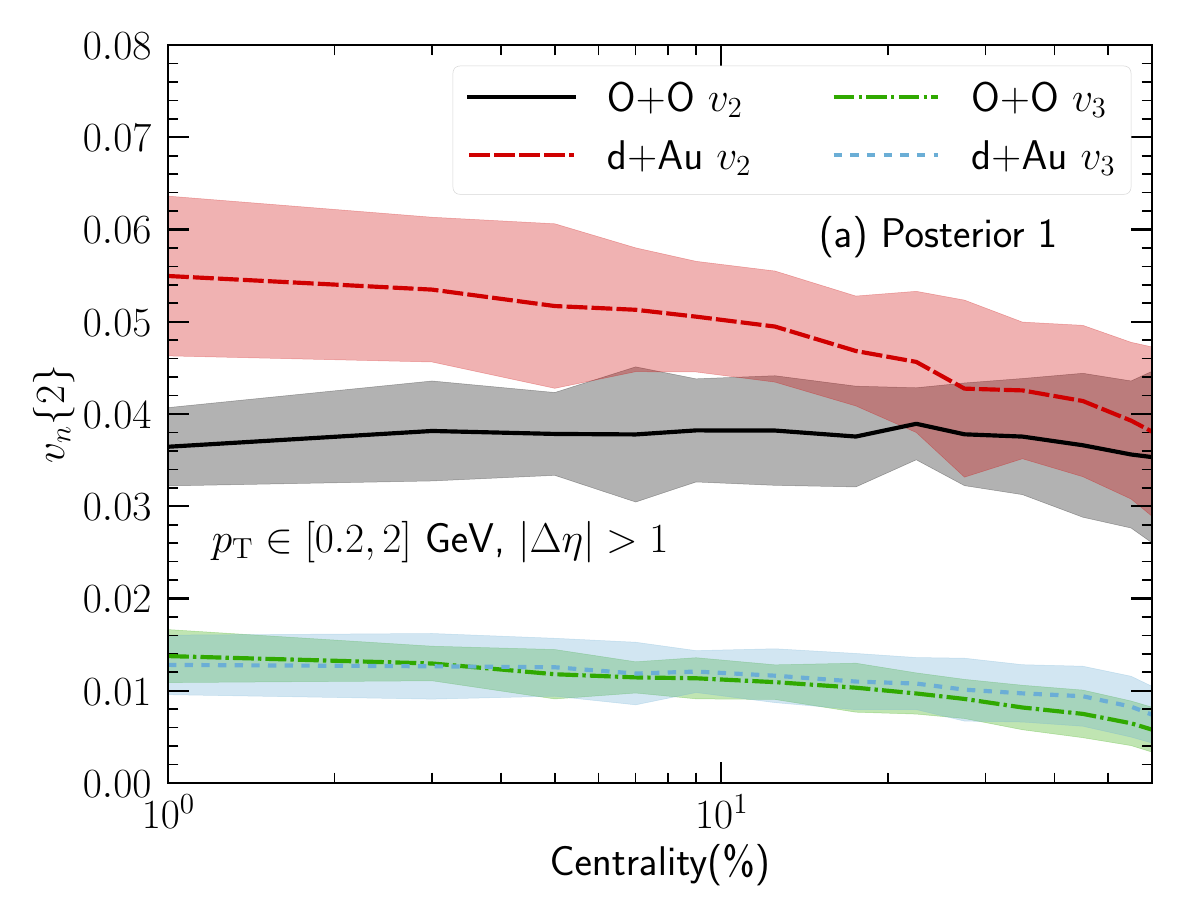}
    \includegraphics[width=\linewidth]{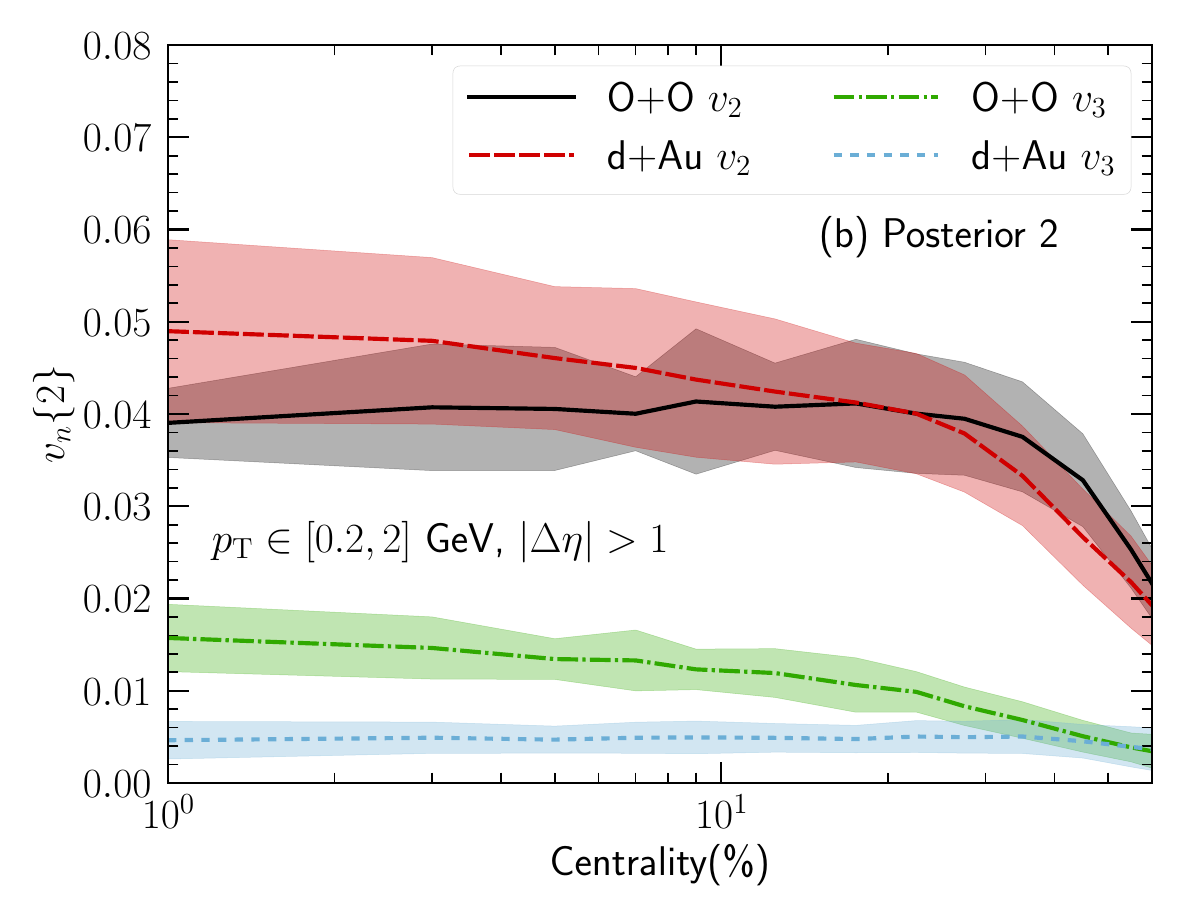}
    \caption{Centrality dependence of anisotropic flow coefficients in O+O and d+Au collisions at $\snn = 200$ GeV using Posterior 1 (a) and 2 (b). The shaded bands represent systematic uncertainty in the theoretical results.}
    \label{fig:PostPred_OOvsdAu}
\end{figure}

Figures~\ref{fig:PostPred_OOvsdAu}a and b show our model predictions using Posterior 1 and 2. 
Our results with Poserior 3 are close to those obtained with Poserior 2. 
We find that the elliptic flow coefficients in central d+Au collisions are 30-50\% larger than those in O+O collisions. 
Our results can be understood as a consequence of the hydrodynamic responses to more eccentric initial geometric shapes in central d+Au collisions compared to central O+O collisions.

Although the Posterior 1 and 2 provide comparable descriptions of $v_n\{2\}$ in the large Au+Au collisions, the anisotropic flow coefficients in small O+O and d+Au collisions show strong sensitivity to the two posterior distributions. 
Elliptic flow coefficients in both collision systems fall off more quickly with the larger values of QGP specific shear viscosity from Posterior 2. 
With model parameters sampled from Posterior 1, we find that the triangular flow coefficients are similar in the two collision systems. 
In contrast, the larger values of the QGP-specific shear viscosity and smaller initial triangularity from Posterior 2 result in smaller $v_3\{2\}$ in d+Au collisions than those in O+O collisions.

\begin{figure}[h!]
    \centering
    \includegraphics[width=\linewidth]{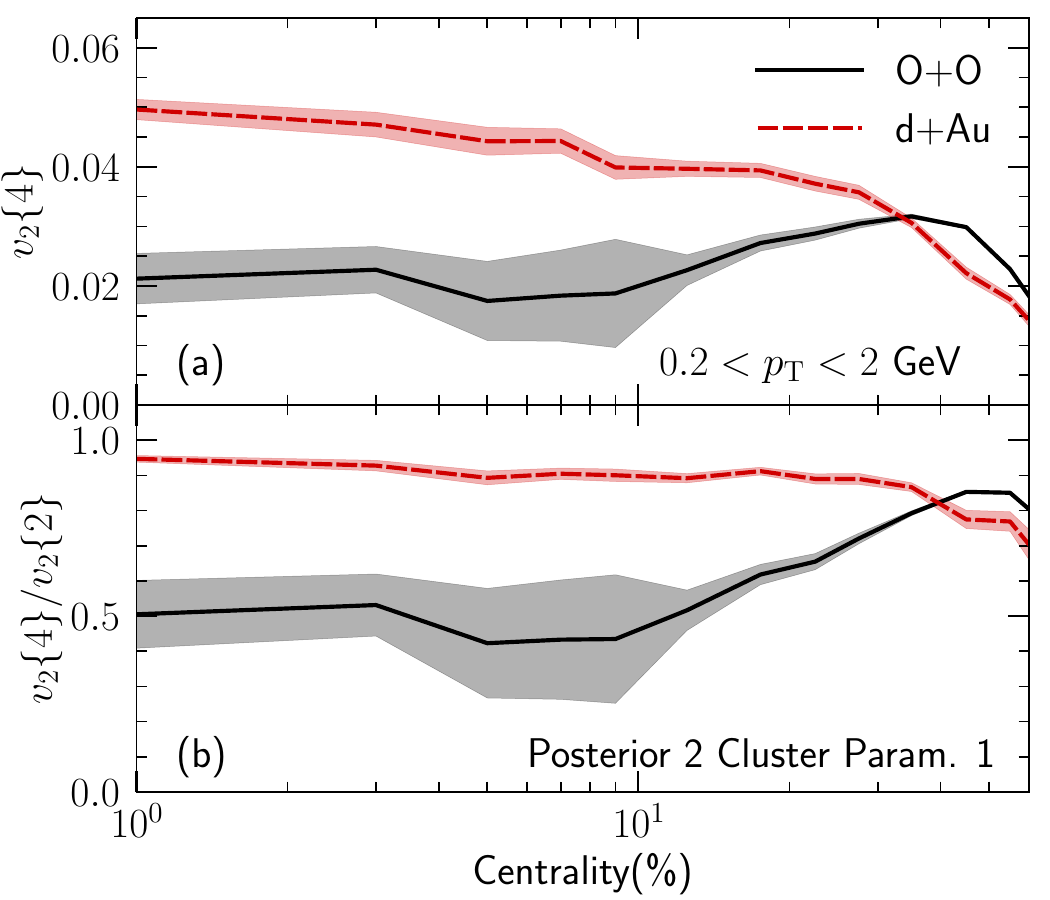}
    \caption{Panel (a): Centrality dependence of the four-particle cumulant elliptic flow $v_2\{4\}$ for O+O and d+Au collisions. Panel (b): The ratios of $v_2\{4\}/v_2\{2\}$ for O+O and d+Au collisions at 200 GeV. The shaded bands represent statistical uncertainty in the theoretical results.}
    \label{fig:PostPred_OOvsdAu_v24}
\end{figure}

Computing anisotropic flow with the four-particle cumulant method requires high statistics. 
Figure~\ref{fig:PostPred_OOvsdAu_v24}a shows the $v_2\{4\}$ in O+O and d+Au collisions with cluster parameter set 1 from the Posterior 2. 
The difference of $v_2\{4\}$ in central d+Au and O+O collisions is twice larger than the difference of $v_2\{2\}$. 
The flow fluctuations give positive contributions to $v_2\{2\}$, while contribute negative to $v_2\{4\}$.
Figure~\ref{fig:PostPred_OOvsdAu_v24}b shows the ratios of $v_2\{4\}/v_2\{2\}$ for O+O and d+Au collisions, which is sensitive to the flow fluctuations. 
This anisotropic flow ratio is about 0.5 in central O+O collisions, while it is around 0.9 for central d+Au collisions. 
It indicates that there are a lot more elliptic flow fluctuations in central O+O collisions than those in d+Au collisions.

To study the hydrodynamic response to initial geometry in these small collision systems, we can compute the mean elliptic flow estimated as~\cite{Borghini:2001vi, Ollitrault:2009ie}
\begin{align}
    \bar{v}_2 = \sqrt{\frac{v_2^2\{2\} + v_2^2\{4\}}{2}}.
\end{align}
Based on our results from Figs.~\ref{fig:PostPred_OOvsdAu} and \ref{fig:PostPred_OOvsdAu_v24}, we have $\bar{v}_2$(O+O)$\approx 0.32$, $\bar{v}_2$(d+Au)$\approx 0.5$, and their ratio $\approx 0.64$ in central collisions.

By correlating anisotropic flow vectors from the STAR TPC and EPD detectors, one can measure the $\pT$-differential anisotropic flow with different rapidity gaps and study the impacts of longitudinal flow decorrelation on these flow observables. 
The $\pT$-differential anisotropic flow with the scalar-product method $v_n\{\mathrm{SP}\}(\pT)$ is defined as
\begin{align}
    v_n\{\mathrm{SP}\}(\pT) = \left\{\begin{array}{ll}
    \frac{\langle V_n(\pT, \eta) Q_n^*(\eta_\mathrm{ref}^A) \rangle}{\sqrt{\langle Q_n(\eta_\mathrm{ref}^A) Q_n^*(\eta_\mathrm{ref}^B)\rangle}} & \mbox{, } \eta \in [0.5, 1.5] \\
    \frac{\langle V_n(\pT, \eta) Q_n^*(\eta_\mathrm{ref}^B) \rangle}{\sqrt{\langle Q_n(\eta_\mathrm{ref}^A) Q_n^*(\eta_\mathrm{ref}^B)\rangle}} & \mbox{, } \eta \in [-1.5, -0.5]
    \end{array} \right. ,
    \label{eq:vnSPpT}
\end{align}
We choose the two sub-events for the reference flow vectors with $-1.5 < \eta_\mathrm{ref}^A < -0.5$ and $0.5 < \eta_\mathrm{ref}^B < 1.5$ from the STAR's TPC detector. 
The TPC flow vectors are computed using charged hadrons with $\pT \in [0.2, 2]$\,GeV.
With these kinematic choices, the $v_2\{\mathrm{SP}\}(\pT)$ with reference flow vectors from the TPC imposes rapidity gaps $|\Delta \eta| > 1$. 

Because the elliptic flow in d+Au collisions is not symmetric in pseudorapidity, we will only use the reference flow vectors from the backward EPD detector in the Au-going side in d+Au collisions. 
In this case, the $\pT$-differential flow is computed with the 3-subevent method,
\begin{align}
    v_n(\pT) = \frac{\langle V_n(\pT) Q^*_n(\eta^A_\mathrm{ref}) \rangle}{\sqrt{\frac{\langle Q_n(\eta^B_\mathrm{ref}) Q^*_n(\eta^A_\mathrm{ref}) \rangle \langle Q_n(\eta^C_\mathrm{ref}) Q^*_n(\eta^A_\mathrm{ref}) \rangle}{\langle Q_n(\eta^B_\mathrm{ref}) Q^*_n(\eta^C_\mathrm{ref}) \rangle}}}.
    \label{eq:vnpTSP3sub}
\end{align}
The reference flow vector $Q_n(\eta^A_\mathrm{ref})$ is computed with the all particles (charged + neutral) in the EPD detector with $\eta^A_\mathrm{ref} \in [-5.1, -2.1]$ and no $\pT$ cuts. The other two reference flow vectors are chosen from the STAR TPC charge tracks with $\pT \in [0.2, 2]$ GeV, $\eta^B_\mathrm{ref} \in [-1.5, -0.5]$, and $\eta^C_\mathrm{ref} \in [0.5, 1.5]$. 
The particles of interest are charged hadrons with $\pT \in [0.2, 2]$ GeV and $\eta \in [-0.5, 0.5]$.

\begin{figure}[h!]
    \centering
    \includegraphics[width=\linewidth]{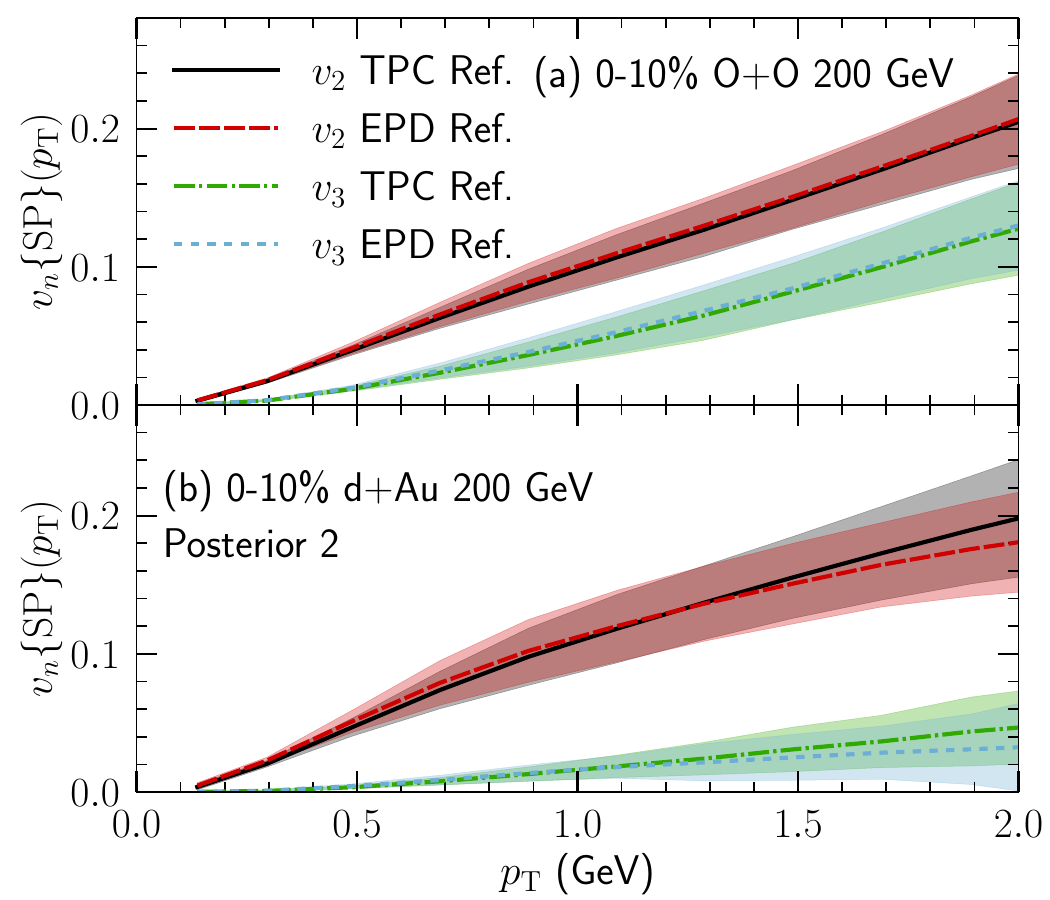}
    \caption{The $\pT$-differential anisotropic flow for 0-10\% O+O (a) and d+Au (b) collisions with two reference anisotropic flow vectors. The shaded bands represent systematic uncertainty in the theoretical results.}
    \label{fig:PostPred_OOvsdAu_vnpT}
\end{figure}

Figures~\ref{fig:PostPred_OOvsdAu_vnpT}a and b show the $\pT$-differential anisotropic flow for 0-10\% O+O and d+Au collisions from Posterior 2. 
Here, we focus on comparing the results using reference flow vectors from the STAR TPC and EPD detectors. 
We do not see a sizable difference between these two sets of observables for both collision systems. 
Naively, one would expect more longitudinal flow decorrelation when using the EPD reference flow vectors, which would result in smaller $v_n\{\mathrm{SP}\}(\pT)$ than those using the reference flow from the TPC. 
The reason for the absence of a noticeable difference in our results is that the rapidity gaps in the denominator of Eq.~\eqref{eq:vnpTSP3sub} are comparable with those in the numerator correlation. They partially cancel with each other, resulting in similar $v_n(\pT)$ from the two reference flow regions.

\begin{figure}[h!]
    \centering
    \includegraphics[width=\linewidth]{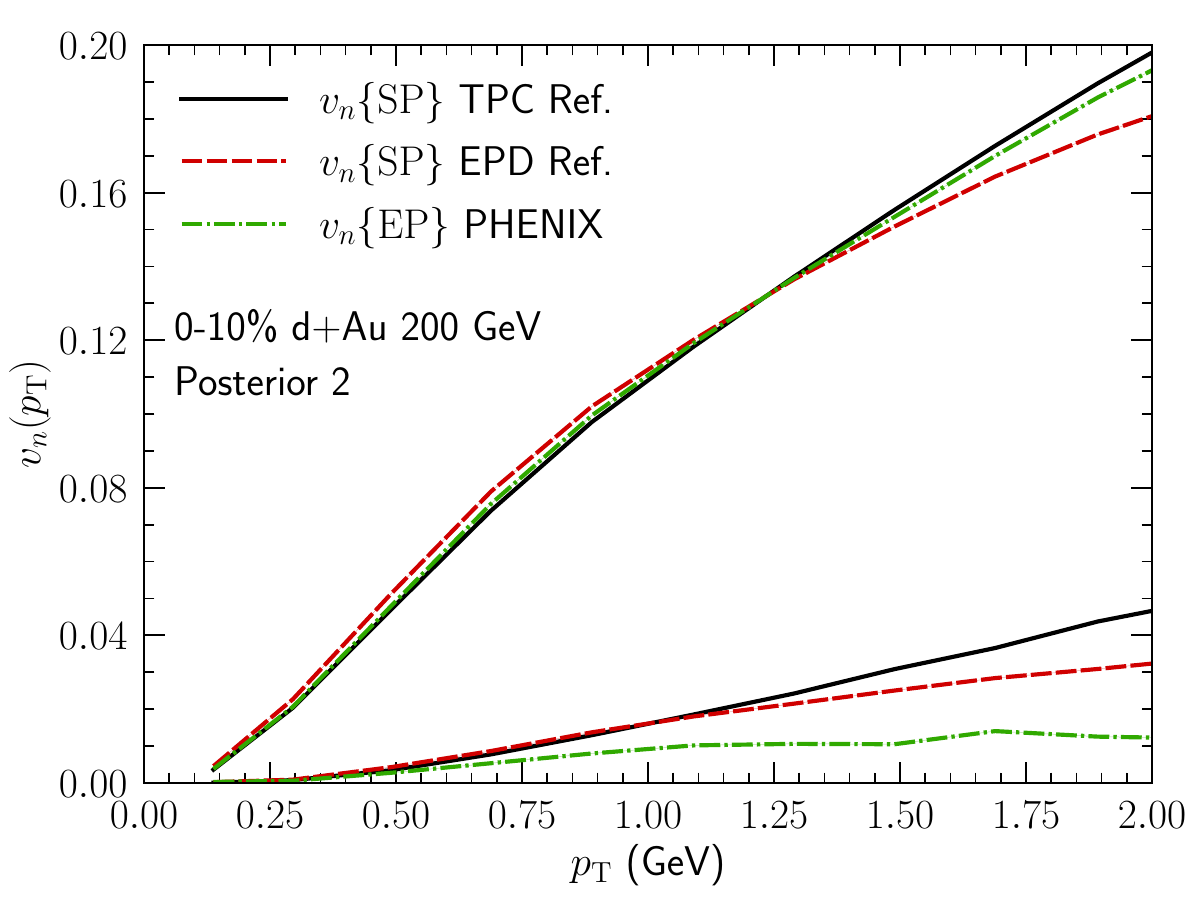}
    \caption{The $\pT$-differential anisotropic flow for 0-10\% d+Au collisions with two reference anisotropic flow vectors compared with results using the PHENIX event-plane method.}
    \label{fig:PostPred_Au_vnpT_STARvsPHENIX}
\end{figure}

Lastly, we compare the $\pT$-differential anisotropic flow with the PHENIX definition using the event-plane method~\cite{PHENIX:2018lia, PHENIX:2018hho}. 
The definition of the PHENIX event-plane method is similar to Eq.~\eqref{eq:vnpTSP3sub}, but replacing all the $Q_n$ vectors by their arguments $Q_n/|Q_n|$. 
The kinematics in the PHENIX detectors are defined as $\eta^A_\mathrm{ref} \in [-3.9, -3.1]$, $\eta^B_\mathrm{ref} \in [-3, -1]$, and $\eta^C_\mathrm{ref} \in [-0.35, 0.35]$~\cite{PHENIX:2018lia, PHENIX:2018hho}. 

For easy comparisons, we show the $v_n(\pT)$ from the three definitions without the systematic uncertainty bands in Fig.~\ref{fig:PostPred_Au_vnpT_STARvsPHENIX}.
We find that the differences in $v_2(\pT)$ among the three different methods are negligible for 0-10\% d+Au collisions. In contrast, the PHENIX $v_3\{\mathrm{EP}\}(\pT)$ is noticeably smaller than the other two methods. The difference in $v_3(\pT)$ in our results is caused by the longitudinal flow decorrelation, which can partially explain the difference in the PHENIX and STAR measurements~\cite{PHENIX:2018lia, STAR:2022pfn, Zhao:2022ugy, Ryu:2023bmx}.

\subsection{Identified particle $v_0(p_T)$ at RHIC BES}

A new observable $v_0(\pT)$ was proposed to probe the fluctuations of hydrodynamic radial flow~\cite{Gardim:2019iah, Schenke:2020uqq}. 
A few studies have shown that it is sensitive to the equation of state, bulk viscosity, and fluctuations at the kinetic freeze-out~\cite{Parida:2024ckk, Gong:2024lhq, Saha:2025nyu}. 
This observable has been recently measured at the LHC energy and agrees well with hydrodynamic predictions~\cite{ALICE:2025iud, ATLAS:2025ztg}. 
Therefore, it is interesting to explore the collision energy dependence of this $\pT$-differential observable in the RHIC BES program. 

\begin{figure}[h!]
    \centering
    \includegraphics[width=\linewidth]{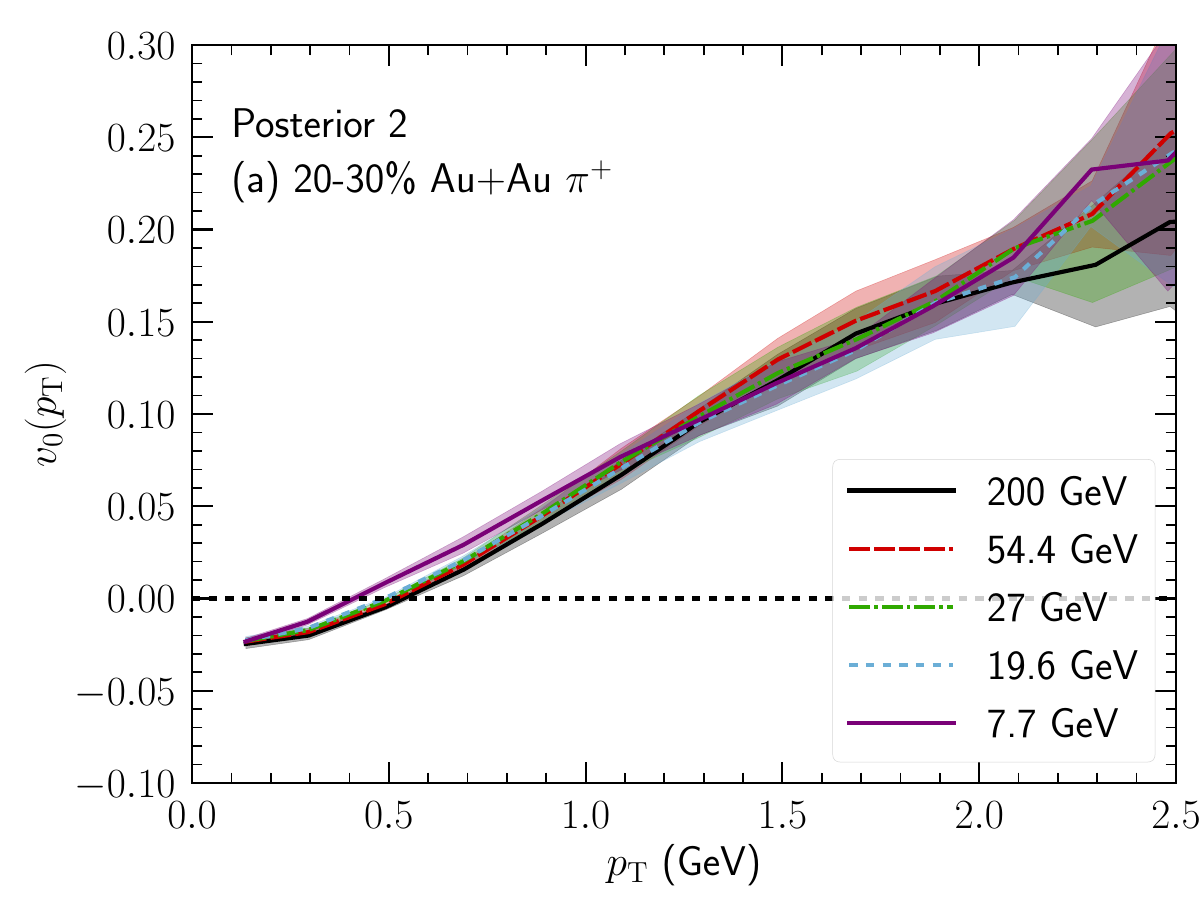}
    \includegraphics[width=\linewidth]{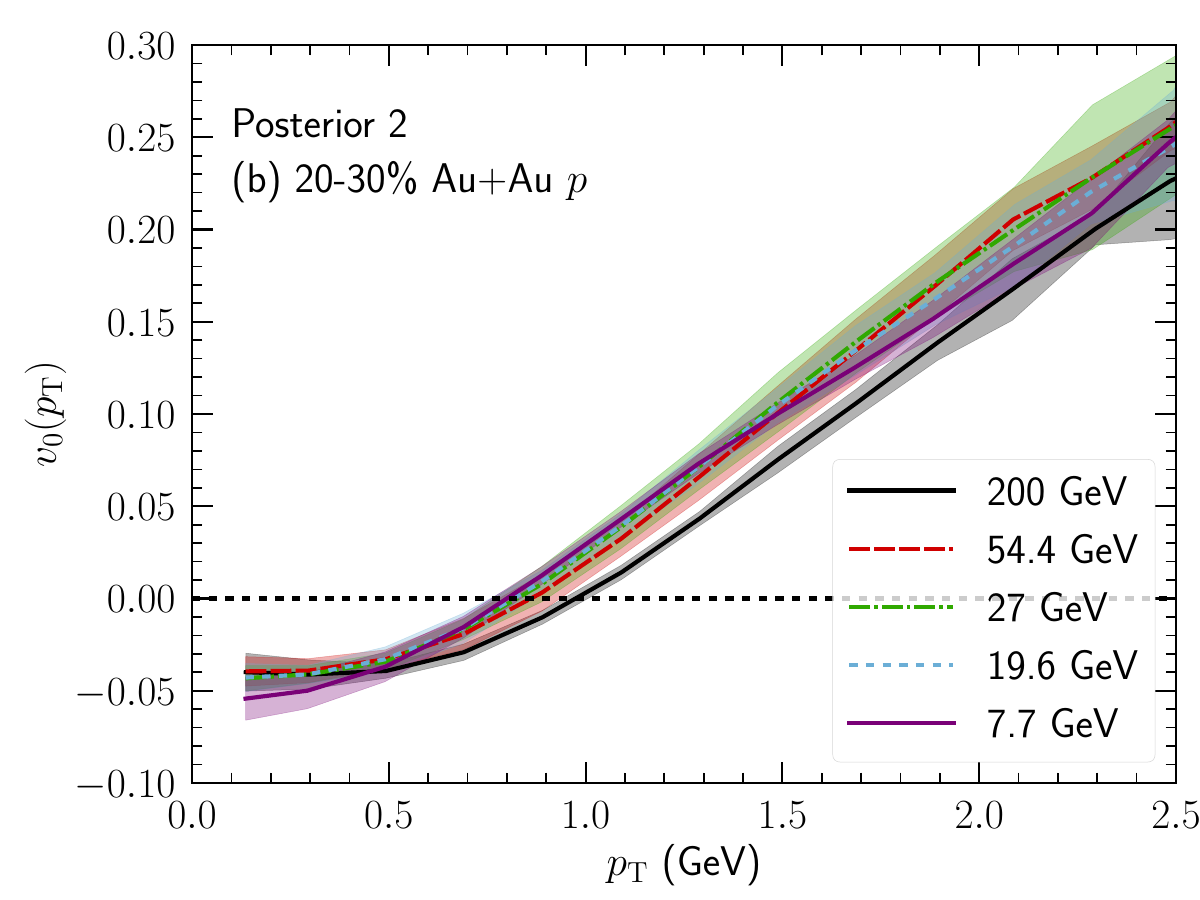}
    \caption{The collision energy dependence of $v_0(\pT)$ for $\pi^+$ and protons in 20-30\% Au+Au collisions in the RHIC BES program. The shaded bands represent systematic uncertainty in the theoretical results.}
    \label{fig:PostPred_v0pT_energydep}
\end{figure}

Figure~\ref{fig:PostPred_v0pT_energydep} shows our model prediction for $v_0(\pT)$ of positively charged pions and protons in 20-30\% Au+Au collisions from 7.7 to 200 GeV. 
We observe similar collision energy dependence for the $v_0(\pT)$ of pions and protons. 
The collision energy dependence is more clearly present in protons' $v_0(\pT)$ because the heavier protons are more sensitive to the hydrodynamic radial flow than pions. 
The shape of the $\pT$-differential $v_0(\pT)$ can be interpreted using the following simple model~\cite{Gardim:2019iah, Schenke:2020uqq},
\begin{align}
    v_0(p_T) \approx \frac{\sigma_{p_T}}{\langle \pT \rangle}\left(2\frac{\pT}{\langle \pT \rangle} - 2\right),
\end{align}
which suggests that $v_0(\pT)$ crosses zero at $\pT = \langle \pT \rangle$ and its slope is proportional to the normalized standard deviation of the $\pT$ fluctuations. 
As the collision energy decreases from 200 to 7.7 GeV, the systems develop less hydrodynamic radial flow, shifting the $v_0(\pT)$ to cross zero at smaller $\pT$ values. 
Meanwhile, radial flow (or $\pT$) fluctuations are also reduced with collision energy, leading to smaller slopes of $v_0(\pT)$, especially for $\snn \le 19.6$\,GeV. 

\begin{figure}[h!]
    \centering
    \includegraphics[width=\linewidth]{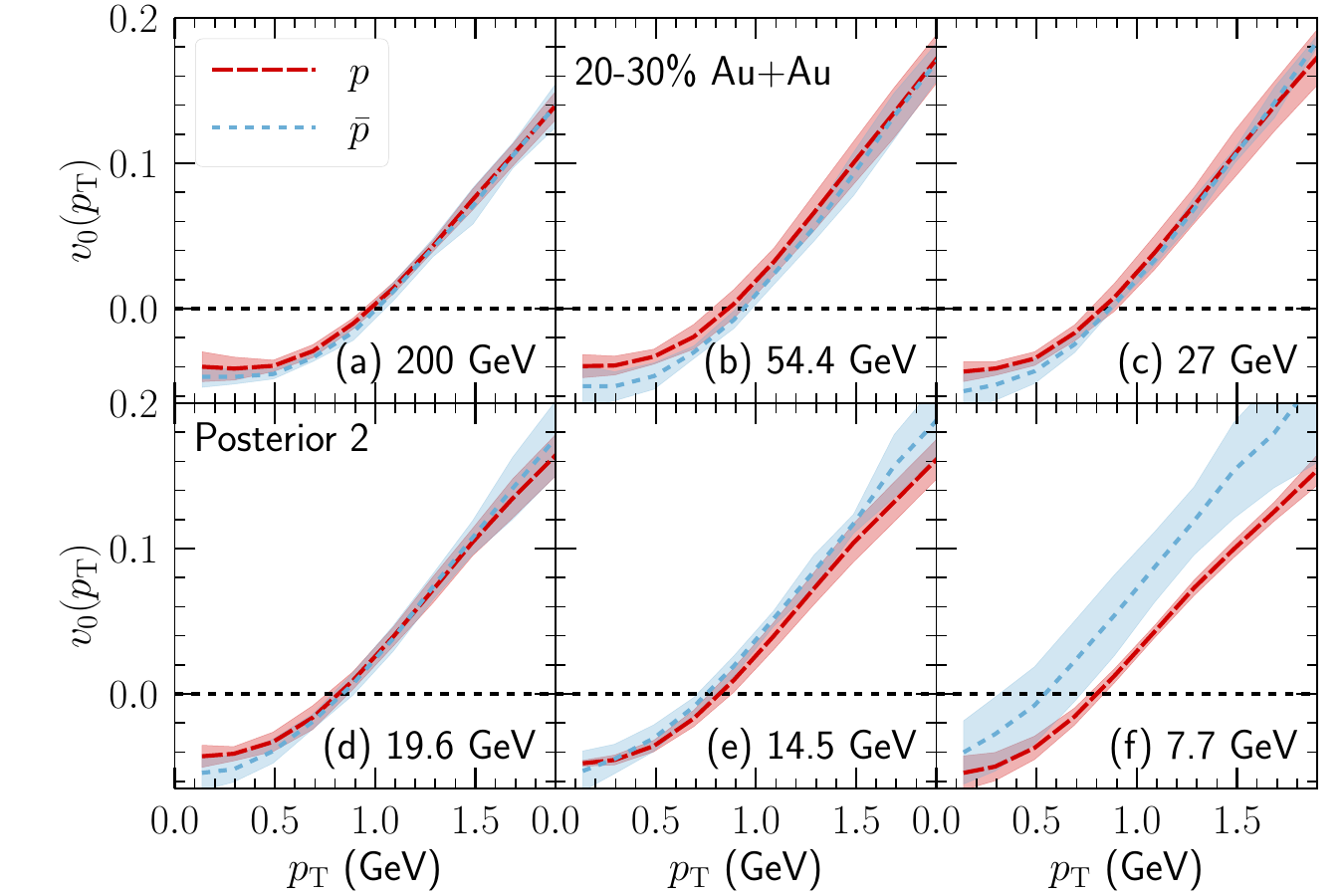}
    \caption{The $v_0(\pT)$ for protons and anti-protons in 20-30\% Au+Au collisions at six energies. The shaded bands represent systematic uncertainty in the theoretical results.}
    \label{fig:PostPred_v0pT_pvspbar}
\end{figure}

Figure~\ref{fig:PostPred_v0pT_pvspbar} shows the comparison of $v_0(\pT)$ between protons and anti-protons in the baryon-rich collision systems at the RHIC BES energies.
Noticeable difference is only observed at 7.7 GeV, where the anti-proton's $v_0(\pT)$ shifts to smaller $\pT$ compared to proton $v_0(\pT)$, because of their mean $\pT$ difference.

It would be interesting to verify these predictions for the $v_0(\pT)$ observables at the RHIC BES program.

\section{Conclusion}
\label{sec:conclusion}

In this work, we have applied the Bayesian model selection method to optimize the (3+1)D hybrid model description of relativistic heavy-ion collisions at the RHIC BES energies.
The Bayes factor provides an unbiased metric for introducing extra model parameters with $\snn$ dependence.
As a result, this method gives only moderate evidence in favor of $\snn$ dependence for the initial hotspot size $\sigma_x$ and $\sigma_\eta$ in our model. 
This result suggests that the \texttt{3D-Glauber+MUSIC+UrQMD} hybrid framework is effective in capturing the collision energy dependence of the QGP properties studied in the RHIC BES program. 
We also use this method to explore whether the current RHIC BES measurements require non-trivial $\mu_B$-dependence of the QGP transport coefficients.

With the optimized model, we study how the model posterior distributions vary when adding more experimental observables, namely all identified particle yields, changed hadron transverse momentum fluctuations, and their $\pT$-differential anisotropic flow. 
This exercise is setting the stage for performing iterative Bayesian inference analyses in the future. 
We observe that the anti-particle to particle yield ratios (hadron chemistry) measured at $\snn \le 19.6$\,GeV favor a low switching energy density $\esw \approx 0.16$\,GeV/fm$^3$.
Because all the model parameters are correlated in the posterior distribution, the low $\esw$ values shift the QGP shear viscosity at $\mu_B = 0$ to larger values to maintain a reasonable description of anisotropic flow measurements in the analysis.

Lastly, we make several model predictions with theoretical uncertainties estimated from the obtained posterior distributions. 
We employ a cluster sampling method to produce a reasonable estimate of theoretical uncertainties using limited samples from the posterior distributions. 
We find that our model posteriors are strongly sensitive to the longitudinal flow decorrelation $r_n(\eta)$ observables, suggesting that precision measurements are valuable for further constraining our (3+1)D model in future Bayesian analyses. 
We compute the pseudorapidity-dependent elliptic flow $v_2(\eta)$ and compare it with existing experimental data.
Our model has some tension to simultaneously describe the collision energy and pseudorapidity of $v_2(\eta)$, indicating it is essential to introduce the temperature and $\mu_B$ dependence of the QGP shear viscosity. 
Our model predictions for the anisotropic flow coefficients in small systems (O+O and d+Au collisions) and the identified particle $v_0(\pT)$ can be compared with upcoming measurements from RHIC to provide additional insights into collectivity in these systems.

\section*{Data availability statement}
We provide the following numerical results of this work for community use:
\begin{itemize}
    \item A snapshot of the \texttt{iEBE-MUSIC} framework for performing event-by-event simulations~\cite{jahan_2025_15920131}
    \item Training data listed in Table~\ref{tab:training_data} and GP emulators~\cite{jahan_2025_15920131}
    \item Posterior chains listed in Table~\ref{tab:posteriorList}~\cite{jahan_2025_15920131}
    \item Bayesian inference Python scripts~\cite{hendrik_roch_2025_15879411}
    \item Simulation data for various model predictions presented in Sec.~\ref{sec:prediction}~\cite{jahan_2025_15920131}
\end{itemize}

\begin{acknowledgments}
We thank Zhenyu Chen, Shengli Huang, Maowu Nie, Takafumi Niida, and Prithwish Tribedy for providing kinematic information related to the upcoming STAR measurements. We thank the members of JETSCAPE SIMS WG for the fruitful discussion.
This work is supported in part by the U.S. Department of Energy, Office of Science, Office of Nuclear Physics, under DOE Award No.~DE-SC0021969 and DE-SC0024232. H.~R. and C.~S. were supported in part by the National Science Foundation (NSF) within the framework of the JETSCAPE collaboration (OAC-2004571).
C.~S. acknowledges a DOE Office of Science Early Career Award. 
S.A.~J. acknowledges support from
Rumble Fellowship and James Kaskas Summer Graduate Scholarship.
Numerical simulations presented in this work were partly performed at the Wayne State Grid, and we gratefully acknowledge their support.
This research was done using computational resources provided by the Open Science Grid (OSG)~\cite{Pordes:2007zzb,Sfiligoi:2009cct,OSPool,OSDF}, which is supported by the National Science Foundation awards \#2030508 and \#2323298.
\end{acknowledgments}

\appendix

\section{Cluster Sampling of Posterior Distributions}
\label{app:posterior_clustering}

\begin{figure*}[t!]
    \includegraphics[width=\textwidth]{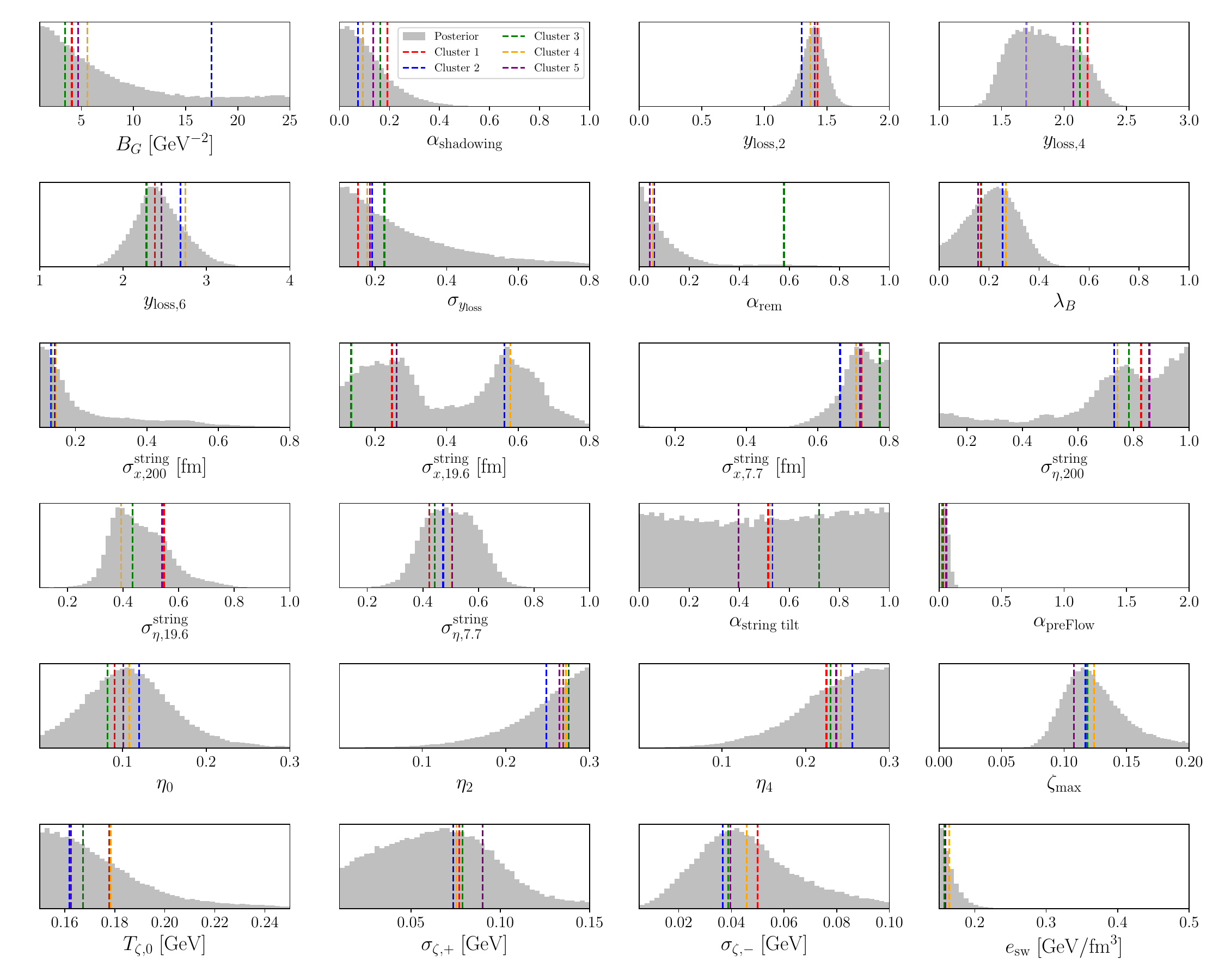}
    \caption{Marginalized posterior distributions from Posterior 2 are shown as gray histograms. Vertical colored lines indicate the positions of the cluster centers identified by the clustering algorithm. The horizontal axes correspond to the ranges of the uniform prior distributions for each parameter, and the vertical axes are in arbitrary units.}
    \label{fig:posterior_cluster_example}
\end{figure*}

In our analysis, we aim to propagate the uncertainty from the inferred posterior distribution to model predictions.
Rather than making predictions solely with the maximum a posteriori (MAP) parameter set, we can sample multiple parameter sets from the posterior distributions and compute the standard deviation of the observables as an estimate for their uncertainties~\cite{Jahan:2024wpj}. 
The uncertainty propagation is crucial for making reliable predictions when the posterior distribution is multimodal or not sharply peaked. 
However, this approach typically requires a substantial amount of computational resources to convolve $\mathcal{O}(100)$ parameter sets from the posterior distribution with high-statistics model simulations~\cite{Jahan:2024wpj}. Therefore, we adopt a clustering-based method to identify a limited number of effective samples with high likelihoods from the posterior distribution~\cite{henderson1982cluster, roesch1993adaptive}. 
We define the model prediction with uncertainty for an observable $O$ as,
\begin{align}
    O = \langle O \rangle_{\theta} \pm \Delta_{\theta} O,
    \label{eq:obsPred}
\end{align}
where
\begin{align}
    \langle O \rangle_{\theta} = \frac{1}{N_\mathrm{samples}} \sum_{\{\bm{\theta}_i\}} O (\bm{\theta}_i)
    \label{eq:eq:obsPred1}
\end{align}
and
\begin{align}
    \Delta_{\theta} O = 2\sqrt{\frac{1}{N_\mathrm{samples}} \sum_{\{\bm{\theta}_i\}} [O (\bm{\theta}_i) -  \langle O \rangle_{\theta}]^2}.
    \label{eq:eq:obsPred2}
\end{align}
Here, we define the theoretical uncertainty as twice the standard deviation computed from simulations with the $N_\mathrm{samples}$ parameter samples because $N_\mathrm{samples} \approx \mathcal{O}(10)$ is a small number. 
By performing high-statistics model simulations on these samples, we can get a reasonable estimate of the model uncertainties $\Delta_{\theta} O$ for the predicted observables with a moderate amount of computational resources.

We start by selecting 1k out of 100k posterior samples with the highest likelihood values (the top 1\%). 
This subset guarantees that we focus on the most probable region of the posterior while maintaining diversity among the samples. 
In the next step, the selected samples are standardized to have a zero mean and unit variance in each parameter dimension, ensuring uniform scaling and preventing parameters with larger numerical ranges from dominating the clustering algorithm.
We then apply the standard K-Means clustering algorithm to the normalized samples, choosing 10 clusters. 
The algorithm separates the samples into 10 disjoint groups by minimizing the variance within the cluster. 
The resulting cluster centers are transformed back to the original parameter space via the inverse of the scaling transformation. 
The produced clusters are interpreted as representative parameter sets that span the distinct regions within the high-probability posterior volume.
A Python script to generate the posterior clusters is part of the Bayesian inference workflow package in Ref.~\cite{hendrik_roch_2025_15879411}.

Figure~\ref{fig:posterior_cluster_example} displays the marginalized posterior distributions obtained from Posterior 2 defined in Table~\ref{tab:posteriorList}.
Vertical colored lines indicate the locations of the first five cluster centers, as determined by the K-Means algorithm~\cite{krishna1999genetic}. 
These cluster centers represent distinct, high-probability regions of the posterior distribution and demonstrate how the clustering-based approach captures posterior structure beyond what is accessible through the MAP estimate alone, especially in cases where the distributions exhibit multimodality.

We perform full model simulations with parameter sets at the individual cluster centers and compute the observables according to Eqs.~\eqref{eq:obsPred}-\eqref{eq:eq:obsPred2}. 
We find the relative standard deviation is about 6\% for $v_2\{2\}$ at 200 GeV. 
To make a comparison, we use the model emulator to compute the same observable at 1,000 parameter samples from the posterior distribution, and we obtain 8\% relative standard deviation in this case. 
This comparison shows that we can get a reasonable estimation of the relative standard deviation for our observable with only five parameter sets from the cluster sampling method. 
To be cautious, we quote $\Delta_\theta O = 2 \sigma$ as the uncertainty for our model predictions in Sec.~\ref{sec:prediction}. 

Tables~\ref{tab:posterior1_clusters}-\ref{tab:posterior3_clusters} list five posterior clusters that we used to generate the results shown in Sec.~\ref {sec:prediction} for the posteriors defined in Tab.~\ref{tab:posteriorList}.

\begin{table}[h!]
    \caption{Posterior 1 cluster parameters.}
    \label{tab:posterior1_clusters}
    \centering
    \begin{tabular}{c|c|c|c|c|c}
        \hline\hline
        Parameter & C1 & C2 & C3 & C4 & C5 \\
        \hline
        $B_G\;[\mathrm{GeV}^{-2}]$ & 16.40 & 17.16 & 15.71 & 13.96 & 15.20 \\
        $\alpha_{\rm shadowing}$ & 0.14 & 0.17 & 0.19 & 0.15 & 0.12 \\
        $y_{{\rm loss},2}$ & 1.49 & 1.54 & 1.43 & 1.40 & 1.38 \\
        $y_{{\rm loss},4}$ & 1.97 & 1.90 & 2.16 & 2.13 & 2.07 \\
        $y_{{\rm loss},6}$ & 2.04 & 2.18 & 1.84 & 1.80 & 1.81 \\
        $\sigma_{y_{\rm loss}}$ & 0.31 & 0.31 & 0.29 & 0.32 & 0.29 \\
        $\alpha_{\rm rem}$ & 0.60 & 0.54 & 0.55 & 0.57 & 0.62 \\
        $\lambda_B$ & 0.10 & 0.18 & 0.19 & 0.18 & 0.16 \\
        $\sigma_{x}^{\rm string}\;[{\rm fm}]$ & 0.11 & 0.11 & 0.12 & 0.11 & 0.11 \\
        $\sigma_{\eta}^{\rm string}$ & 0.29 & 0.19 & 0.18 & 0.18 & 0.19 \\
        $\alpha_{{\rm string}\;{\rm tilt}}$ & 0.67 & 0.72 & 0.62 & 0.75 & 0.66 \\
        $\alpha_{\rm preFlow}$ & 0.04 & 0.06 & 0.06 & 0.16 & 0.05 \\
        $\eta_0$ & 0.04 & 0.05 & 0.07 & 0.06 & 0.04 \\
        $\eta_2$ & 0.26 & 0.24 & 0.26 & 0.27 & 0.26 \\
        $\eta_4$ & 0.27 & 0.26 & 0.28 & 0.28 & 0.27 \\
        $\zeta_{\rm max}$ & 0.14 & 0.14 & 0.15 & 0.15 & 0.14 \\
        $T_{\zeta,0}\;[{\rm GeV}]$ & 0.20 & 0.20 & 0.19 & 0.18 & 0.18 \\
        $\sigma_{\zeta,+}\;[{\rm GeV}]$ & 0.03 & 0.02 & 0.02 & 0.03 & 0.04 \\
        $\sigma_{\zeta,-}\;[{\rm GeV}]$ & 0.03 & 0.04 & 0.03 & 0.02 & 0.02 \\
        $e_{\rm sw}\;[{\rm GeV}/{\rm fm}^3]$ & 0.38 & 0.33 & 0.33 & 0.33 & 0.34 \\
        \hline\hline
    \end{tabular}
\end{table}

\begin{table}[h!]
    \caption{Posterior 2 cluster parameters.}
    \label{tab:posterior2_clusters}
    \centering
    \begin{tabular}{c|c|c|c|c|c}
        \hline\hline
        Parameter & C1 & C2 & C3 & C4 & C5 \\
        \hline
        $B_G\;[\mathrm{GeV}^{-2}]$ & 4.09 & 17.48 & 3.45 & 5.59 & 4.69 \\
        $\alpha_{\rm shadowing}$ & 0.19 & 0.07 & 0.16 & 0.09 & 0.14 \\
        $y_{{\rm loss},2}$ & 1.43 & 1.30 & 1.37 & 1.37 & 1.40 \\
        $y_{{\rm loss},4}$ & 2.19 & 1.70 & 2.12 & 1.70 & 2.07 \\
        $y_{{\rm loss},6}$ & 2.38 & 2.69 & 2.28 & 2.74 & 2.46 \\
        $\sigma_{y_{\rm loss}}$ & 0.15 & 0.19 & 0.23 & 0.18 & 0.19 \\
        $\alpha_{\rm rem}$ & 0.06 & 0.06 & 0.58 & 0.05 & 0.04 \\
        $\lambda_B$ & 0.17 & 0.25 & 0.17 & 0.27 & 0.16 \\
        $\sigma_{x,200}^{\rm string}\;[{\rm fm}]$ & 0.14 & 0.13 & 0.13 & 0.14 & 0.14 \\
        $\sigma_{x,19.6}^{\rm string}\;[{\rm fm}]$ & 0.25 & 0.56 & 0.13 & 0.58 & 0.26 \\
        $\sigma_{x,7.7}^{\rm string}\;[{\rm fm}]$ & 0.72 & 0.66 & 0.77 & 0.71 & 0.72 \\
        $\sigma_{\eta,200}^{\rm string}$ & 0.83 & 0.73 & 0.78 & 0.74 & 0.86 \\
        $\sigma_{\eta,19.6}^{\rm string}$ & 0.55 & 0.39 & 0.43 & 0.39 & 0.54 \\
        $\sigma_{\eta,7.7}^{\rm string}$ & 0.42 & 0.47 & 0.44 & 0.50 & 0.51 \\
        $\alpha_{{\rm string}\;{\rm tilt}}$ & 0.52 & 0.53 & 0.72 & 0.53 & 0.40 \\
        $\alpha_{\rm preFlow}$ & 0.03 & 0.06 & 0.02 & 0.06 & 0.05 \\
        $\eta_0$ & 0.09 & 0.12 & 0.08 & 0.11 & 0.10 \\
        $\eta_2$ & 0.27 & 0.25 & 0.28 & 0.27 & 0.26 \\
        $\eta_4$ & 0.22 & 0.26 & 0.23 & 0.24 & 0.24 \\
        $\zeta_{\rm max}$ & 0.12 & 0.12 & 0.12 & 0.12 & 0.11 \\
        $T_{\zeta,0}\;[{\rm GeV}]$ & 0.18 & 0.16 & 0.18 & 0.18 & 0.16 \\
        $\sigma_{\zeta,+}\;[{\rm GeV}]$ & 0.08 & 0.07 & 0.08 & 0.08 & 0.09 \\
        $\sigma_{\zeta,-}\;[{\rm GeV}]$ & 0.05 & 0.04 & 0.04 & 0.05 & 0.04 \\
        $e_{\rm sw}\;[{\rm GeV}/{\rm fm}^3]$ & 0.16 & 0.16 & 0.16 & 0.16 & 0.16 \\
        \hline\hline
    \end{tabular}
\end{table}

\begin{table}[h!]
    \caption{Posterior 3 cluster parameters.}
    \label{tab:posterior3_clusters}
    \centering
    \begin{tabular}{c|c|c|c|c|c}
        \hline\hline
        Parameter & C1 & C2 & C3 & C4 & C5 \\
        \hline
        $B_G\;[\mathrm{GeV}^{-2}]$ & 3.93 & 2.50 & 2.64 & 2.40 & 2.27 \\
        $\alpha_{\rm shadowing}$ & 0.23 & 0.20 & 0.23 & 0.20 & 0.23 \\
        $y_{{\rm loss},2}$ & 1.44 & 1.50 & 1.39 & 1.42 & 1.46 \\
        $y_{{\rm loss},4}$ & 2.14 & 2.14 & 2.15 & 2.18 & 2.14 \\
        $y_{{\rm loss},6}$ & 2.28 & 2.34 & 2.26 & 2.31 & 2.30 \\
        $\sigma_{y_{\rm loss}}$ & 0.13 & 0.14 & 0.14 & 0.14 & 0.16 \\
        $\alpha_{\rm rem}$ & 0.07 & 0.06 & 0.08 & 0.09 & 0.08 \\
        $\lambda_B$ & 0.20 & 0.23 & 0.24 & 0.20 & 0.24 \\
        $\sigma_{x,200}^{\rm string}\;[{\rm fm}]$ & 0.15 & 0.15 & 0.15 & 0.15 & 0.15 \\
        $\sigma_{x,19.6}^{\rm string}\;[{\rm fm}]$ & 0.25 & 0.25 & 0.27 & 0.26 & 0.27 \\
        $\sigma_{x,7.7}^{\rm string}\;[{\rm fm}]$ & 0.76 & 0.79 & 0.78 & 0.79 & 0.78 \\
        $\sigma_{\eta,200}^{\rm string}$ & 0.76 & 0.76 & 0.77 & 0.84 & 0.76 \\
        $\sigma_{\eta,19.6}^{\rm string}$ & 0.51 & 0.52 & 0.50 & 0.51 & 0.51 \\
        $\sigma_{\eta,7.7}^{\rm string}$ & 0.44 & 0.50 & 0.46 & 0.48 & 0.45 \\
        $\alpha_{{\rm string}\;{\rm tilt}}$ & 0.55 & 0.66 & 0.62 & 0.50 & 0.61 \\
        $\alpha_{\rm preFlow}$ & 0.01 & 0.02 & 0.02 & 0.02 & 0.02 \\
        $\eta_0$ & 0.15 & 0.13 & 0.16 & 0.13 & 0.16 \\
        $\eta_2$ & 0.27 & 0.27 & 0.27 & 0.28 & 0.25 \\
        $\eta_4$ & 0.13 & 0.09 & 0.16 & 0.11 & 0.14 \\
        $\zeta_{\rm max}$ & 0.11 & 0.12 & 0.13 & 0.11 & 0.11 \\
        $T_{\zeta,0}\;[{\rm GeV}]$ & 0.16 & 0.19 & 0.16 & 0.16 & 0.17 \\
        $\sigma_{\zeta,+}\;[{\rm GeV}]$ & 0.09 & 0.08 & 0.05 & 0.09 & 0.09 \\
        $\sigma_{\zeta,-}\;[{\rm GeV}]$ & 0.04 & 0.06 & 0.03 & 0.04 & 0.06 \\
        $e_{\rm sw}\;[{\rm GeV}/{\rm fm}^3]$ & 0.16 & 0.17 & 0.16 & 0.17 & 0.16 \\
        \hline\hline
    \end{tabular}
\end{table}
Notably, although the use of cluster sampling algorithms is popular in data analysis, our current work explores how to adopt this technique to sample parameter sets from the posterior distribution for efficient model predictions.

\section{Comparisons of different $\delta f$ corrections at particlization}
\label{Sec:deltaf}

The out-of-equilibrium ($\delta f$) corrections are introduced at the Cooper-Frye particlization procedure as,
\begin{align}
    E \frac{dN_i}{d^3p} = \frac{g_i}{(2\pi)^3} \int_\Sigma d \sigma_\mu p^\mu (f_{i, \mathrm{eq}} + \delta f_i).
    \label{eq:CooperFyre}
\end{align}
where $g_i$ is the spin degeneracy factor, $\Sigma$ is the hydrodynamic hyper-surface with a constant energy density $\esw$ and $d\sigma_\mu$ are the normal vector of the surface elements. 
The $\delta f$ corrections ensure the shear stress tensor $\pi^{\mu\nu}$ and bulk viscous pressure $\Pi$ are mapped properly to the particle's momentum distributions.
\begin{align}
    \sum_i \frac{g_i}{(2\pi)^3} \int \frac{d^3 p}{E} p^\mu p^\nu \delta f_i = \pi^{\mu\nu} - \Pi \Delta^{\mu\nu},
    \label{eq:deltafMactchingCondition}
\end{align}
where the left-hand side sums over all hadronic particle species $i$ and $\Delta^{\mu\nu} = g^{\mu\nu} - u^\mu u^\nu$.
In general, mapping the out-of-equilibrium corrections to individual particle distributions from macroscopic hydrodynamic variables, as in Eq.~\eqref{eq:deltafMactchingCondition}, does not have a unique solution~\cite{Molnar:2014fva}. 
In the literature, Grad's Moment~\cite{Teaney:2003kp, Dusling:2009df, Denicol:2012cn} and Chapman-Enskog (CE)~\cite{Anderson:1974nyl, Jaiswal:2014isa, Czajka:2017wdo, Czajka:2020mho} methods have been used extensively as the $\delta f$ corrections.

At finite densities, we consider three types of conserved charges, namely net baryon ($B$), net electric charges ($Q$), and net strangeness ($S$). 
For each conserved charge, we have one matching condition
\begin{align}
    \sum_i \frac{Q_i g_i}{(2\pi)^3} \int \frac{d^3 p}{E} p^\mu \delta f_i = V_{Q_i}^\mu, \label{eq:deltafMatching_Q_i}
\end{align}
where $Q_i = B, Q, S$ are the quantum charges for each hadron species and $V_{Q_i}^\mu$ is the charge diffusion current in the system~\cite{Denicol:2018wdp}. 
Here, we ignore the charge diffusion currents in our simulations, $V_{Q_i}^\mu = 0$.

The Grad's moment method assumes the $\delta f$ of particle species $i$ takes the form
\begin{align}
    \delta f_i^\mathrm{Grad} & = f_{i, \mathrm{eq}} (1 \pm f_{i, \mathrm{eq}}) \nonumber \\
    & \hspace{0.5cm} [p^\alpha p^\beta \epsilon_{\alpha \beta} + p^\alpha (B_i \epsilon_{B, \alpha} + Q_i \epsilon_{Q, \alpha} + S_i \epsilon_{S, \alpha})],
    \label{eq:Grad_deltaf}
\end{align}
where the coefficients $\epsilon_{\alpha \beta}$, $\epsilon_{B, \alpha}$, $\epsilon_{Q, \alpha}$, $\epsilon_{S, \alpha}$ depend on thermodynamic integrals and can be evaluated with the matching conditions Eqs.~\eqref{eq:deltafMactchingCondition} and \eqref{eq:deltafMatching_Q_i}. 
They are functions of $T$, $\mu_B$, $\mu_Q$, $\mu_S$. 
Eq.~\eqref{eq:Grad_deltaf} is the generalized Grad's $(10 + 4 N_Q)$-moment method at finite densities with $N_Q$ flavors of conserved charges in the system.

The Chapman-Enskog $\delta f$ at finite densities takes the following form for the shear and bulk sectors,
\begin{align}
    \delta f^\mathrm{CE}_{i, \mathrm{shear}} & = f_{i, \mathrm{eq}} (1 \pm f_{i, \mathrm{eq}}) \left( \frac{\pi_{\mu\nu}}{2 \hat{\eta}} \right) \frac{p^{\mu} p^{\nu}}{(p \cdot u) T} \label{eq:CE_shear_df} \\
    \delta f^\mathrm{CE}_{i, \mathrm{bulk}} & = f_{i, \mathrm{eq}} (1 \pm f_{i, \mathrm{eq}}) \left( \frac{\Pi}{\hat{\zeta}} \right) \nonumber \\
    & \hspace{0.5cm} \times \frac{1}{(p \cdot u) T} \left[ \frac{1}{3} m^2 - \left( \frac{1}{3} - \hat{c}^2 \right) (p \cdot u)^2 \right], \label{eq:CE_bulk_df}
\end{align}
where the coefficients $\hat{\eta}$, $\hat{\zeta}$, and $\hat{c}^2$ are thermodynamic integrals computed with the matching conditions as in Eqs.~\eqref{eq:deltafMactchingCondition} and \eqref{eq:deltafMatching_Q_i}. 
They are 4-dimensional functions of $T$, $\mu_B$, $\mu_Q$, $\mu_S$.

The $\delta f$ corrections have been identified as one of the significant theoretical uncertainties in the hybrid modeling of relativistic heavy-ion collisions~\cite{JETSCAPE:2020mzn}. 
A comparative analysis using Bayesian methods was performed at zero net baryon density, and moderate evidence was obtained in favor of Grad's moment method compared to the CE method, as the CE method was worse at fitting the ratios of proton to pion yields.

\begin{figure}[h!]
    \centering
    \includegraphics[width=\linewidth]{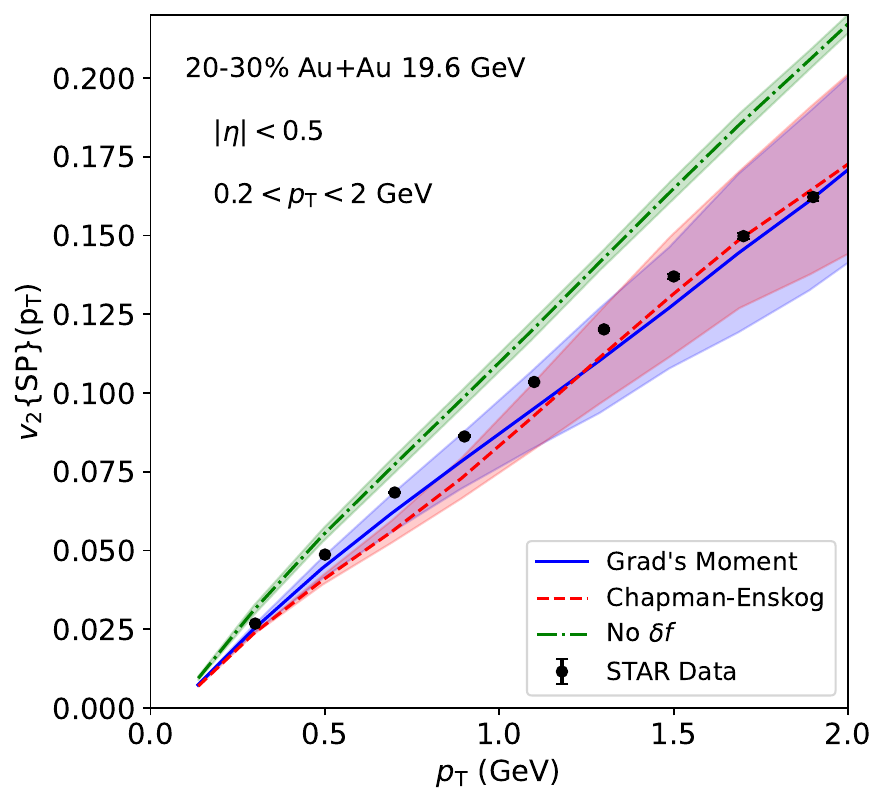}
    \caption{The scalar-product $v_{2}\{\mathrm{SP}\}(\pT)$ for charged hadrons in 20-30\% Au+Au collisions at 19.6 GeV using Poseterior 2. The shaded bands represent systematic uncertainty in the theoretical results. Results using different $\delta f$ corrections, namely Grad's method, Chapman-Enskog method, and without $\delta f$ corrections, are compared to STAR measurements~\cite{STAR:2012och}.}
    \label{fig:deltaf_v2pT}
\end{figure}

In Fig.~\ref{fig:deltaf_v2pT}, we compare the charged hadron $\pT$-differential elliptic flow $v_{2}\{\mathrm{SP}\}(\pT)$ for three cases: 1) no viscous correction (labelled as ``no $\delta f$''), 2) Grad's moment method, and 3) Chapman-Enskog method. 
The differences of $v_{2}\{\mathrm{SP}\}(\pT)$ to the results without $\delta f$ corrections show the sizes of out-of-equilibrium corrections to elliptic flow as a function of $\pT$, which reaches about 30\% at $\pT$ around 2 GeV. 
The model results using the Grad's moment method and the CE approximation are comparable with each other at high $\pT$. 
When $\pT \in [0.5, 1]$\,GeV, the elliptic flow from the CE method is slightly smaller than that with the Grad's moment $\delta f$. 

\begin{figure}[h!]
    \centering
    \includegraphics[width=\linewidth]{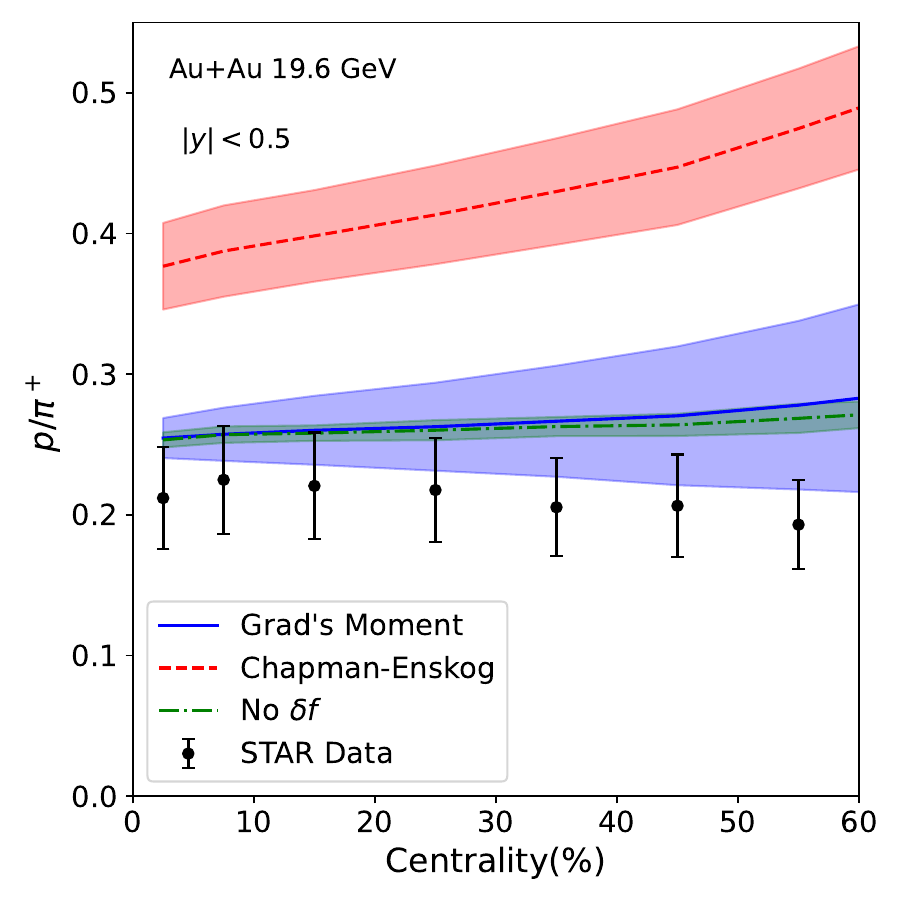}
    \caption{The $p/\pi^{+}$ yield ratio as a function of centrality for 19.6 GeV Au+Au collisions. Results using different $\delta f$ corrections are compared to STAR experimental data~\cite{STAR:2017sal}. The shaded bands represent systematic uncertainty in the theoretical results.
    }
    \label{fig:deltaf_pidRatio}
\end{figure}

More significant differences between the two types of out-of-equilibrium corrections are observed in the ratios of $p/\pi^+$ yields as shown in Fig.~\ref{fig:deltaf_pidRatio}. 
In the ``no $\delta f$'' case, the $p/\pi^+$ ratios are controlled by the chemical freeze-out condition. 
In our hybrid model, the hadron chemistry starts to evolve out-of-equilibrium at the particlization surface with the constant energy density $\esw$ when fluid cells are converted to particles and further propagate within the hadronic transport model. 
With the choice of $\esw \approx 0.16$\,GeV/fm$^3$ from Posterior 2, our model can describe the STAR measurements in central to semi-peripheral Au+Au collisions. 
The results using the Grad's moment $\delta f$ stay close to the ``no $\delta f$'' case; meanwhile, the theoretical uncertainty increases significantly. 
This is because the out-of-equilibrium corrections from bulk viscosity introduce sizable corrections to $p/\pi^+$ ratios when $\esw$ is fixed (see Table~\ref{tab:posterior2_clusters}). 
The lower part of the uncertainty band of Grad's moment $\delta f$ has a reasonable agreement with the STAR data from central up to 60\% centrality. 
In contrast, the results from the Chapman-Enskog method significantly overpredict the $p/\pi^+$ ratio, which could lead to tension when performing Bayesian inference analysis with the CE $\delta f$. 
Figure~\ref{fig:deltaf_pidRatio} demonstrates that the hadronic chemistry is a sensitive probe to constrain the out-of-equilibrium corrections in the particlization procedure.

\section{Model-to-data comparison for the Bayesian Calibration}
\label{app:BayesianCalibration}

\begin{figure*}[h!]
    \includegraphics[width=0.95\textwidth]{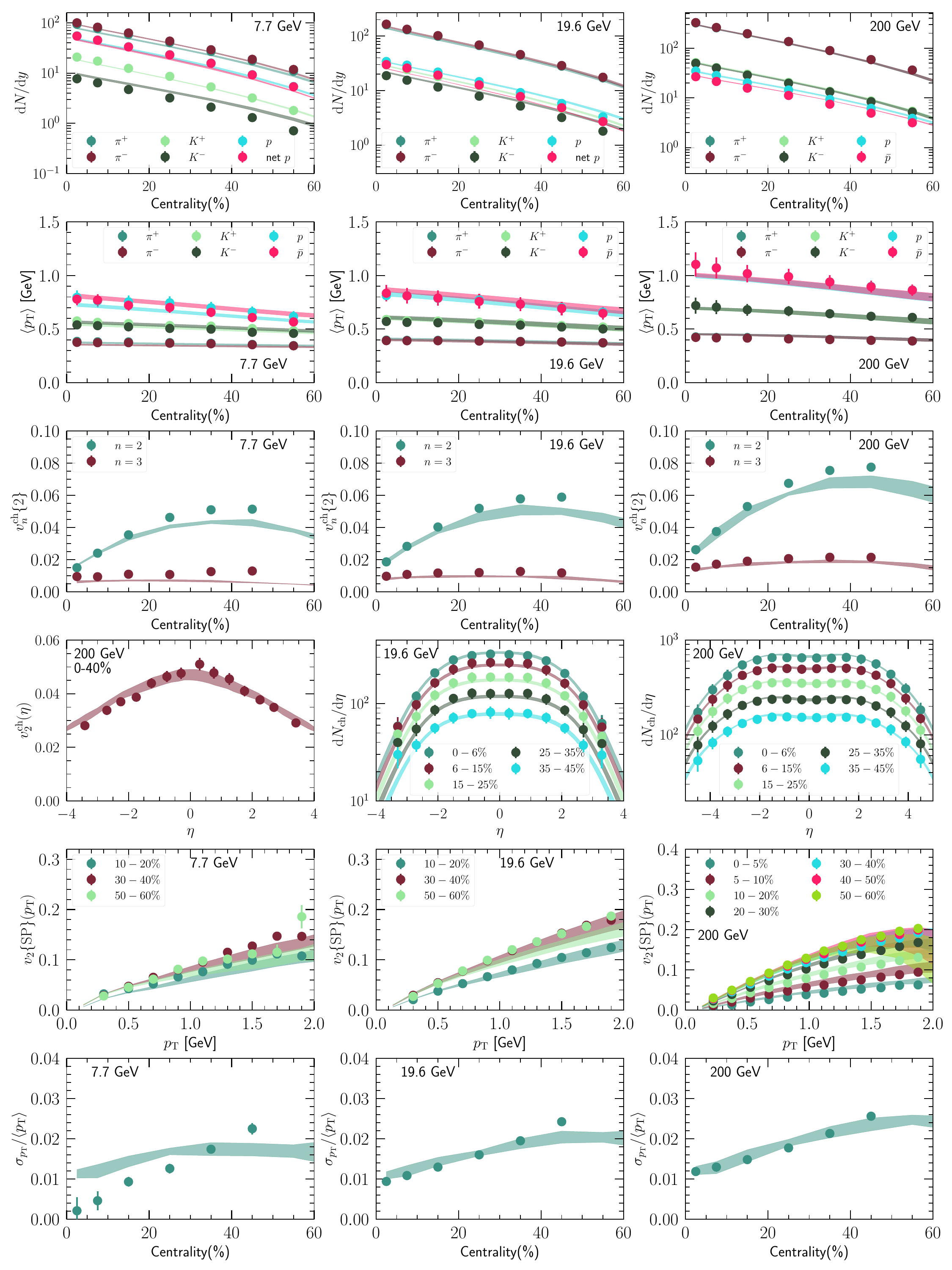}
    \caption{The model simulation results with Posterior 1 compared with experimental measurements listed in Table~\ref{tab:training_data}. The shaded bands represent systematic uncertainty in the theoretical results.}
    \label{fig:ModelCalibrationPost1}
\end{figure*}

\begin{figure*}[h!]
    \includegraphics[width=0.95\textwidth]{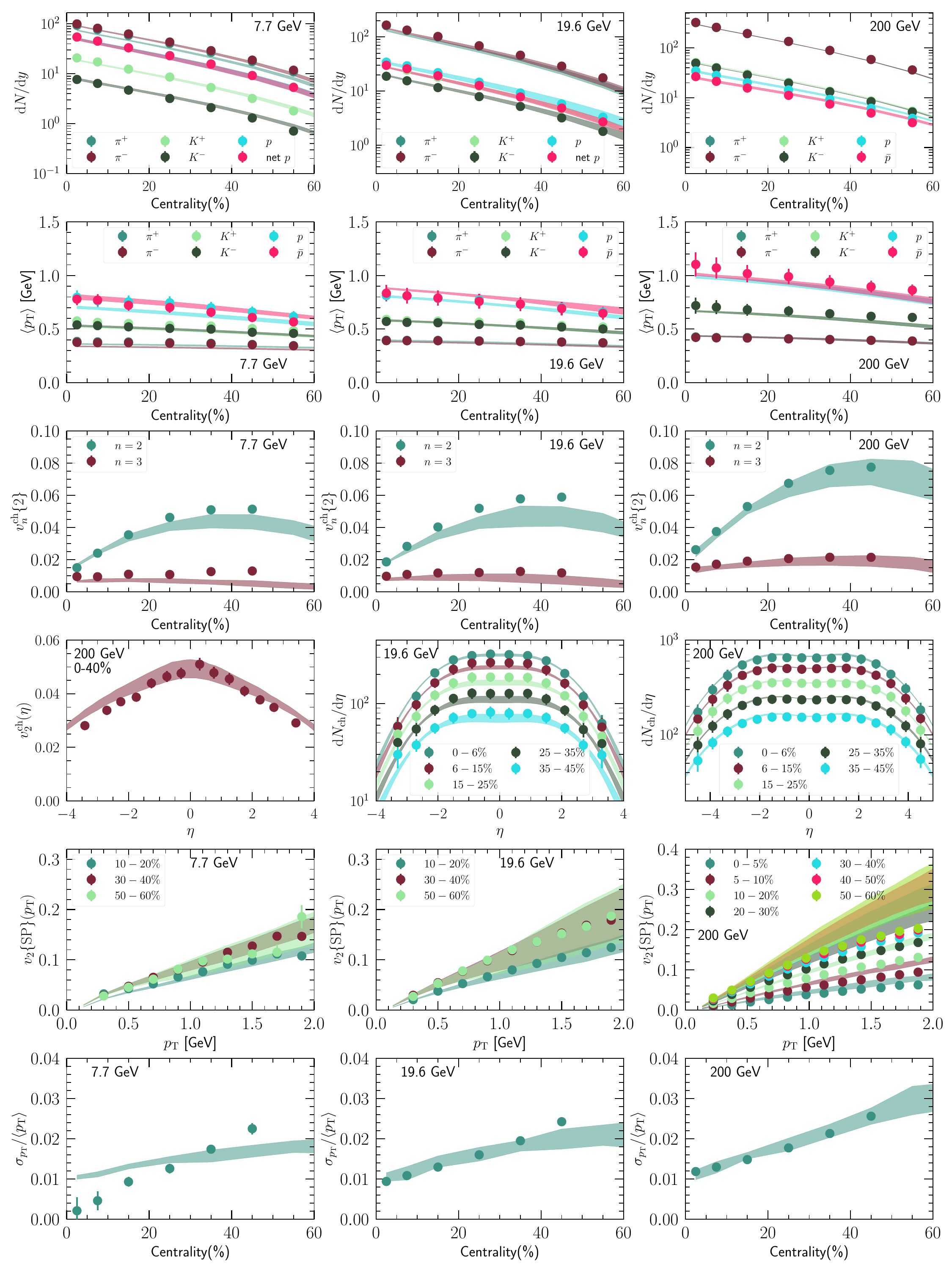}
    \caption{Similar to Fig.~\ref{fig:ModelCalibrationPost1} but for model simulation results with Posterior 2.}
    \label{fig:ModelCalibrationPost2}
\end{figure*}

\begin{figure*}[h!]
    \includegraphics[width=0.95\textwidth]{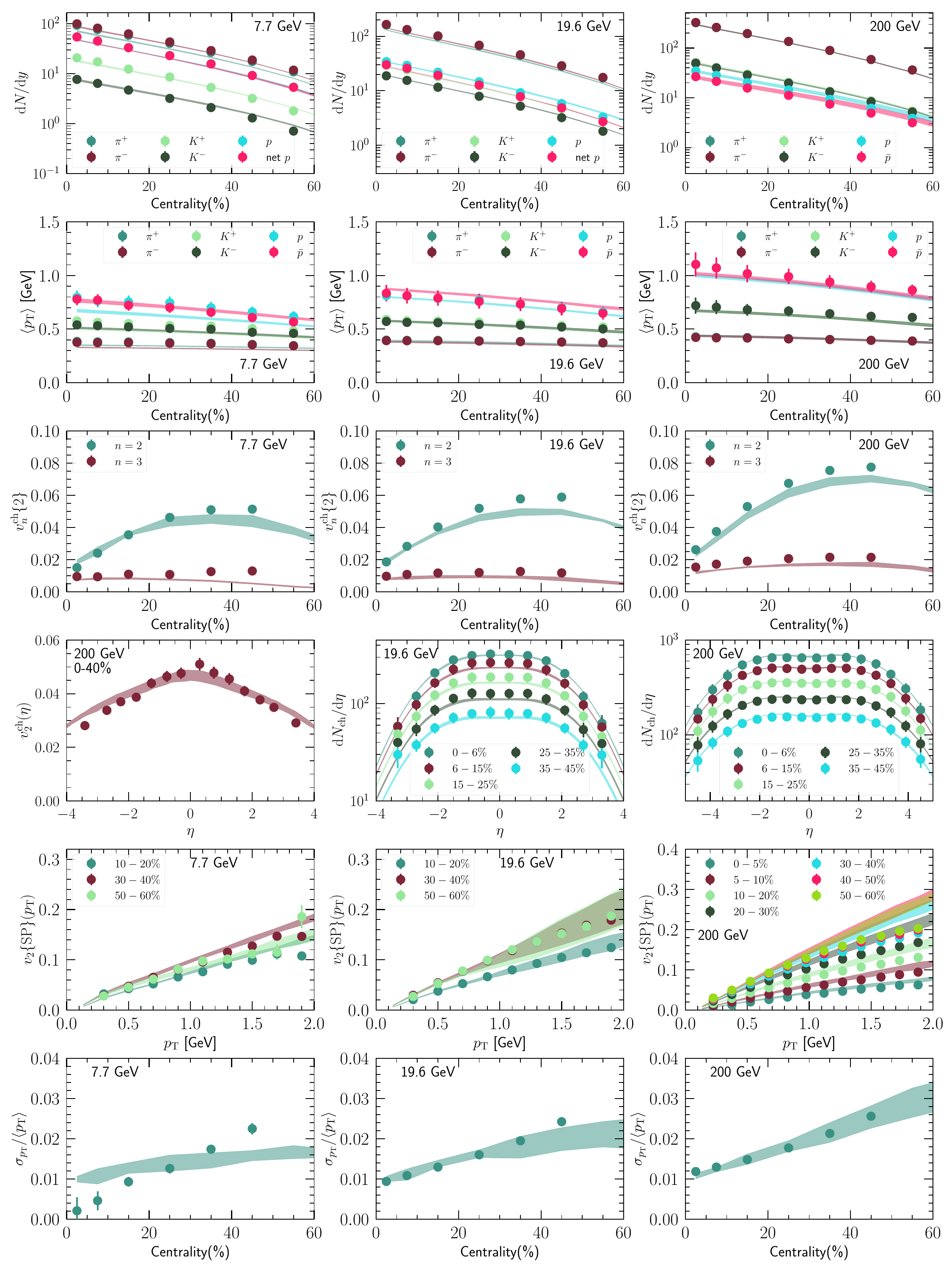}
    \caption{Similar to Fig.~\ref{fig:ModelCalibrationPost1} but for model simulation results with Posterior 3.}
    \label{fig:ModelCalibrationPost3}
\end{figure*}

In this appendix, we provide gallery figures (Figs.~\ref{fig:ModelCalibrationPost1}-\ref{fig:ModelCalibrationPost3}) for our model descriptions of the calibrated data listed in Table~\ref{tab:training_data} from the three posterior distributions.

We observe that our model has failed to capture the centrality dependence of charged hadron $\pT$ fluctuations in Au+Au collisions at 7.7 GeV. 
In central collisions, the STAR measurements drop close to zero, while our model gives $\sigma_{\pT}/\langle \pT \rangle \approx 0.01$, similar to their values at the other two higher collision energies.

\clearpage
\newpage
\bibliography{bib, non-inspire}

\end{document}